\documentclass[aps,prd,preprint,superscriptaddress,tightenlines,nofootinbib]{revtex4}

\usepackage{graphicx}% Include figure files
\usepackage{dcolumn}% Align table columns on decimal point
\usepackage{bm}% bold math

\begin{document}

\preprint{CLEO CONF 06-2}   % For conference papers
\preprint{CLNS 05/1927}   
\preprint{CLEO 05-15}   

\title{Measurement of the Direct Photon Momentum Spectrum in 
$\Upsilon({\rm 1S})$, 
$\Upsilon({\rm 2S})$, and 
$\Upsilon({\rm 3S})$ Decays}
\thanks{Submitted to the 33$^{\rm rd}$ International Conference on High Energy
Physics, July 26 - August 2, 2006, Moscow}

\author{D.~Besson}
\author{S.~Henderson}
\affiliation{University of Kansas, Lawrence, Kansas 66045}
\author{T.~K.~Pedlar}
\affiliation{Luther College, Decorah, Iowa 52101}
\author{D.~Cronin-Hennessy}
\author{K.~Y.~Gao}
\author{D.~T.~Gong}
\author{J.~Hietala}
\author{Y.~Kubota}
\author{T.~Klein}
\author{B.~W.~Lang}
\author{S.~Z.~Li}
\author{R.~Poling}
\author{A.~W.~Scott}
\author{A.~Smith}
\affiliation{University of Minnesota, Minneapolis, Minnesota 55455}
% ...
\collaboration{CLEO Collaboration} %FOR PRL,CLNS
\noaffiliation

\noaffiliation

%-------- END INSERT ------------

%please hard code the date when you have a final draft and submit to CLEOAC
\date{July 22, 2006}

\begin{abstract} 
Using data taken with the CLEO III detector at the Cornell Electron Storage
Ring, we have investigated the direct photon spectrum in the decays
${\Upsilon}({\rm 1S}) \to \gamma gg$,
${\Upsilon}({\rm 2S}) \to \gamma gg$,
${\Upsilon}({\rm 3S}) \to \gamma gg$. The latter two of these are first 
measurements.
Our analysis procedures differ from previous ones in the
following ways: a) background estimates (primarily
from $\pi^0$ decays) are based on isospin symmetry rather
than a determination of the $\pi^0$ spectrum, which permits 
measurement of the $\Upsilon$(2S) and $\Upsilon$(3S) direct photon spectra
without explicit corrections for $\pi^0$ backgrounds from, e.g.,
$\chi_{bJ}$ states,
b) we estimate the branching fractions with
a parametrized functional form (exponential) used for the background,
c) we use the high-statistics sample of
$\Upsilon$(2S)$\to\pi\pi\Upsilon$(1S) to obtain a tagged-sample of
$\Upsilon$(1S)$\to\gamma+X$ events, for which there are no
QED backgrounds. 
We determine values for
the ratio of the inclusive direct photon
decay rate to that of the dominant 
three-gluon decay $\Upsilon \to ggg$ ($R_\gamma=B(gg\gamma)/B(ggg)$) to
be: $R_\gamma$(1S)=(2.70$\pm$0.01$\pm$0.13$\pm$0.24)\%,
$R_\gamma$(2S)=(3.18$\pm$0.04$\pm$0.22$\pm$0.41)\%, and
$R_\gamma$(3S)=(2.72$\pm$0.06$\pm$0.32$\pm$0.37)\%, 
where the errors shown are
statistical, systematic, and theoretical 
model-dependent, respectively.
Given a value of $Q^2$, one can estimate a
value for the strong coupling constant $\alpha_s(Q^2)$ from $R_\gamma$.
\end{abstract}

\pacs{13.20.-v, 13.40.Hq, 14.40.Gx}
\maketitle

\section*{Introduction}

Production of a $B{\overline B}$ meson pair, the
Zweig-favored decay mode of $\Upsilon$ 
mesons, is not energetically possible
for resonances below the $\Upsilon$(4S), thus
the decay 
of the $\Upsilon$(1S) meson must proceed through
Zweig-suppressed channels.
Since the charge conjugation 
quantum number of the $\Upsilon$ resonances is C=--1,
the three lowest-order hadronic decay modes
of the $\Upsilon$(1S) meson are those into three gluons ($ggg$), 
the vacuum
polarization QED decay $\Upsilon\to q{\overline{q}}$, 
and two gluons plus single photon ($gg\gamma$). 
For the $\Upsilon$(2S)
and $\Upsilon$(3S)
resonances, direct radiative transitions, both electromagnetic and
hadronic, compete with these annihilation modes.
Since
$\Gamma_{ggg} \propto \alpha_s^3$ and $\Gamma_{gg\gamma} \propto  
\alpha_s^2\alpha_{em},$
the ratio of the decay rates from these two
processes can be expressed in terms of the
strong coupling constant\cite{r:Brod-Lep-Mack}:
\begin{equation}
R_{\gamma} \equiv \frac{\Gamma_{gg\gamma}}{\Gamma_{ggg}} = 
                  \frac{N_{gg\gamma}}{N_{ggg}}           =
	\frac{38}{5}q_b^2\frac{\alpha_{em}}{\alpha_s}[1+(2.2\pm0.8)\alpha_s/\pi].
\end{equation}
In this expression, the bottom quark charge $q_b=-1/3$.
Alternately, one can normalize to the well-measured 
dimuon channel\cite{r:danko}
and cancel the electromagnetic vertex:
$(\Gamma_{gg\gamma}/\Gamma_{\mu\mu}) \propto \alpha_s^2$.

In either case, one must define the value of $Q^2$ appropriate for 
this process. Although the value $Q^2\sim M_\Upsilon^2$ seems `natural',
the original prescription of Brodsky 
{\it et al.}\cite{r:Brod-Lep-Mack} gave $Q^2=(0.157M_{\Upsilon({\rm 1S})})^2$
for $\Upsilon$(1S)$\to gg\gamma$.

Theory prescribes
the differential spectrum $d^2N/dx_\gamma d\cos\theta_z$ 
($x_\gamma=p_\gamma/E_{\mathrm {beam}}$, and
$\cos\theta_z$ is 
defined as the polar angle relative to the $e^+e^-$ beam axis).
The limited angular coverage of the high-resolution CLEO~III
photon detection, as well
as the large backgrounds at low momentum due to decays of
neutral hadrons (primarily $\pi^0$, $\eta$, $\eta'$
and $\omega$) to photons,
plus the large number of radiated
final state `fragmentation' photons
in this regime limit our sensitivity to the
region defined by $|\cos\theta_z|<$0.7 and $x_\gamma>$0.4.
We must therefore rely on models for
comparison with the observed direct photon
spectrum and extrapolation of the
direct photon spectrum (excluding the
fragmentation component) into lower-momentum
and larger polar angle regions. 

Originally, the decay of the ground-state vector
$b{\overline b}$ bottomonium into three 
vectors (both $\Upsilon\to ggg$ as well as $\Upsilon\to gg\gamma$)
was modeled in lowest-order
QCD after similar QED 
decays of orthopositronium 
into three photons, leading to the expectation that
the direct photon spectrum should rise linearly with $x_\gamma$ 
to the kinematic
limit ($x_\gamma\to 1$); 
phase space considerations lead to a slight enhancement 
exactly at the kinematic limit\cite{r:lowest-qcd,r:kol-walsh}. 
Koller and Walsh considered the angular
spectrum in detail\cite{r:kol-walsh}, demonstrating that as the
momentum of the
most energetic primary parton (photon or gluon)
in $\Upsilon\to\gamma gg$ or $\Upsilon\to ggg$ 
approaches the beam energy, the event axis tends
to align with the beam axis: 
$x_\gamma\to 1\Rightarrow dN/d(\cos\theta_z)\to 1+\cos^2\theta_z$. 
Field\cite{r:Field} argued that
$x_\gamma=1$ is non-physical, since it corresponds to a recoil $gg$ system
with zero invariant mass, while the recoil system must have enough
mass to produce on-shell final state hadrons. Using a phenomenological
parton shower Monte Carlo technique
which took into account the correlation of photon momentum with recoil
hadronization phase space, Field predicted a significant softening of the
lowest-order QCD predicted spectrum, with a photon momentum
distribution peaking at $x_\gamma\sim 0.65$ rather than $x_\gamma\to 1$. 
(In the limit of completely independent fragmentation, the same argument,
in principle, would apply to 3-gluon decays.)
This result seemed in conflict with the extant CUSB\cite{r:CUSB84} data, which 
indicated a spectrum more similar to 
the lowest-order QCD prediction. A subsequent measurement by
CLEO-I\cite{r:Csorna}, 
however, favored Field's softened spectrum over lowest-order
QCD. Given the poor resolution of the
CLEO-I electromagnetic calorimeter,
that measurement was also consistent with 
a subsequent modification to lowest-order QCD which
calculated corrections at the endpoint\cite{r:Photiadis} by
summing leading logs of the form ln(1-$x_\gamma$). Higher
statistics measurements by Crystal Ball\cite{r:Bizeti} 
as well as ARGUS\cite{r:Albrecht-87} corroborated this
softened photon spectrum.\footnote{It is important to note here that all
these measurements assumed that the Koller-Walsh angular distribution
was still applicable to the phenomenological Field model.}
A subsequent CLEO analysis (CLEO-II)\cite{r:nedpaper}, based on
$\sim$1~M $\Upsilon$(1S) events, provided a high-statistics confirmation
of a photon spectrum peaking at $x_\gamma\sim 0.65$, and was able to 
trace the direct photon momentum spectrum down to $x_\gamma\approx$0.4;
at that momentum, the direct photon signal becomes
less than 10\% relative to the background, whereas the systematic errors on
the background estimate 
in that momentum region exceed 10\%.\footnote{We emphasize
here that the CLEO-II analysis, in presenting the 
background-subtracted direct photon spectrum, showed only statistical errors,
whereas the systematic errors in the region $x_\gamma<0.4$ 
are considerably larger
than those statistical errors. This is a point that was not made strongly
enough in the past, encouraging various theoretical fragmentation models
to be tested against those direct photon data at low $x_\gamma$ values.}
Contemporary with the CLEO-II analysis, 
Catani and Hautmann first pointed out complications due to the
presence of
fragmentation photons emitted from final--state
light quarks downstream of the initial heavy quarkonia decay\cite{r:Catani}
(essentially final state radiation). 
These can dominate the
background-subtracted spectra for $x_\gamma<$0.4
and therefore (if not 
corrected for) lead to an over-estimate of the 
$\Upsilon$(1S)$\to gg\gamma$ branching fraction, and an underestimate
of the extracted value of $\alpha_s$.

Hoodbhoy and Yusuf\cite{r:YusufHoodbhoy96} also performed a rigorous
calculation of the expected $\Upsilon$(1S)$\to gg\gamma$ decay rate, by
summing all the diagrams contributing to the direct photon final state
and treating hard and soft contributions separately.
Rather than assuming that the decay occurs via annihilation of two
at-rest quarks, the authors smear the annihilation over a size
of order $1/m$, with a corresponding non-zero velocity.
Although their
calculation results in some softening of the photon spectrum relative to the
lowest-order QCD prediction, it is unable to entirely account for the
softening observed in data, leading to the conclusion that 
final-state gluon interactions are
important, particularly near the photon endpoint.

Fleming and Leibovich\cite{r:Fleming03a} 
considered the photon spectrum in three distinct
momentum regions. At low momentum ($x_\gamma<0.3$), final-state
radiation effects dominate.
In the intermediate momentum regime (0.3$<x_\gamma<$0.7), they applied 
the operator product expansion
(OPE) to the direct photon spectrum of $\Upsilon$ decay, with
power-counting rules prescribed by non-relativistic QCD (NRQCD), 
and retained
only the lowest-order color singlet terms in $v/c$. In the highest-momentum
regime ($x_\gamma>0.7$), a soft-collinear effective theory (SCET) for the
light degrees of freedom 
combined with non-relativistic QCD for the
heavy degrees of freedom was used to obtain a prediction for the 
photon spectrum which qualitatively described the essential features of the
CLEO-II data, despite peaking at a higher value of $x_\gamma$ than data.
The same approach was later applied
by Fleming to decays of the type $e^+e^-\to J/\psi$+X, given the
similarity to $e^+e^-\to\Upsilon$(1S)$\to\gamma$+X, and including the
color-octet contributions to $J/\psi$ production\cite{FLM03}.

Very recently, Garcia and Soto (GS\cite{r:SotoGarcia04}) 
have also produced a parameterization of
the expected photon momentum spectrum in the $\Upsilon$ system. 
Following Fleming and Leibovich, they also
remedy the inability of non-relativistic QCD (NRQCD)
to model the
endpoint region by combining NRQCD with Soft-Collinear Effective Theory,
which allows calculation of the spectrum of the collinear gluons
resulting as $x_\gamma\to 1$.
They make their own calculation of the octet contributions (in both
S- and P- partial waves) to the overall rate, obtaining a spectral shape
prediction similar to Fleming and Leibovich (claimed to be reliable in
the interval 0.65$\le x_\gamma\le$0.92\cite{r:GarciaPrivateComm}) 
after adding color-octet, color-singlet,
and fragmentation contributions. For $x_\gamma\le$0.65, 
Garcia-Soto consider the fragmentation
contribution ``significant'' compared to the direct photon
spectrum. For $x_\gamma\ge$0.92, the calculation becomes less reliable;
in this high-momentum regime, the possibility of two-body decays:
$\Upsilon\to gg\gamma\to {\cal X}\gamma$, with ${\cal X}$ some
resonant hadronic state, will also lead to distortions of the 
expected spectrum. Contributions from such possible two-body
decays may also result in a slight underestimate of the
extracted value of $\alpha_s$. Garcia and Soto have also pointed
out the possibility of different calculational regimes for
$\Upsilon$(1S)$\to\gamma+X$, compared to
$\Upsilon$(2S)$\to\gamma+X$ and
$\Upsilon$(3S)$\to\gamma+X$, given the difference in the principal quantum
numbers, and therefore the average radial interquark separation. 
Since the Field model is based on simple gluon-gluon
fragmentation phase space arguments, it does not distinguish 
between direct photons from any of the three $\Upsilon$ resonances.

\section*{Overview of Analysis}
\begin{figure}[htpb]
\centerline{\includegraphics[width=8cm]{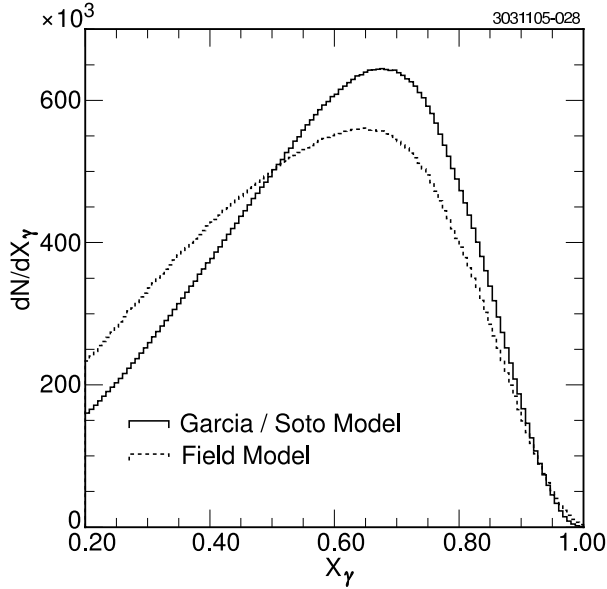}}
\caption{Comparison of the direct photon spectral shapes for the
two theoretical models used in this analysis.}
\label{fig:FieldGS}
\end{figure}

The analysis, in general terms, proceeds as follows. After selecting a
high-quality sample of $e^+e^-$ annihilations into
hadrons, we plot the inclusive isolated
photon spectrum in data taken
at both on-resonance and off-resonance energies. A direct
subtraction of the off-resonance contribution isolates the photon
spectrum due to $\Upsilon$ decays. The background from decays of neutral
hadrons into
photons ($\pi^0\to\gamma\gamma$,
$\eta\to\gamma\gamma$, $\eta'\to\rho\gamma$, 
and $\omega\to\pi^0\gamma$) produced in $\Upsilon$ decays
to $ggg$, $gg\gamma$, or $q{\overline q}$
is removed statistically using a Monte Carlo generator
developed specifically for this purpose, and based on the assumption
that the kinematics of charged and neutral hadron production can
be related through isospin
conservation. For each charged pion identified in the
data, we simulate a two-body decay of one of the neutral hadrons enumerated
above. The measured four-momentum of that charged pion is then used
to boost the daughter photons into the lab frame. After correcting for
efficiency, and scaling by the expected rate of neutral hadron
production relative to charged pion production (for $\pi^0$'s, the
simple isospin assumption would be $N(\pi^0)/N(\pi^\pm)\sim$1/2; for
the other neutral hadrons, we use ratios relative to charged
pions as derived in our previous analysis\cite{r:nedpaper}) 
and the appropriate branching fractions, 
a background ``pseudo-photon'' spectrum is created. 
After subtracting all backgrounds, the
remaining photon spectrum is interpreted as the direct photon spectrum, 
which must then be extrapolated into low-photon momentum and high
$\cos\theta_z$ regions (for which
the backgrounds are prohibitively large) 
in order to determine an estimate of the
full production rate.
In this analysis, we 
employ the models by Field and Garcia-Soto
for integration purposes, given their acceptable
match in spectral shape to
previous data. Although no predictions exist for direct photon
decays of the $\Upsilon$(2S) and $\Upsilon$(3S) resonances, we nevertheless
use these same models to determine total direct photon decay rates in the
case of these higher resonances. A comparison of the shapes of these models is 
shown in Figure \ref{fig:FieldGS}.

\section*{\label{sect:criterion}Data Sets and Event Criteria}
The CLEO~III detector is a general purpose solenoidal magnet spectrometer and
calorimeter. Elements of the detector, as well as performance characteristics,
are described in detail elsewhere \cite{r:CLEO-II,r:CLEOIIIa,r:CLEOIIIb}. 
For photons in the central
``barrel'' region of the cesium iodide (CsI)
electromagnetic calorimeter, at energies
greater than 2 GeV, the energy resolution
is given by
\begin{equation}
\frac{ \sigma_{\rm E}}{E}(\%) = \frac{0.6}{E^{0.7309}} + 1.14 - 0.01E,
                                \label{eq:resolution1}
\end{equation}
where $E$ is the shower energy in GeV. At 100 MeV, the
calorimetric performance is about 20\% poorer than indicated by this expression
due to the material in front of the calorimeter itself. The tracking system, 
Ring Imaging Cerenkov Detector (RICH) particle identification system, 
and electromagnetic
calorimeter are all contained within a 1 Tesla superconducting
coil.

\section*{Event Selection}
The data used in this analysis were collected on the $\Upsilon$(1S) resonance,
center-of-mass energy $E_{\mathrm CM} = 9.46$ GeV, the $\Upsilon$(2S) resonance, 
center-of-mass energy $E_{\mathrm CM} = 10.02$ GeV, and the $\Upsilon$(3S) resonance, 
center-of-mass energy $E_{\mathrm CM} = 10.36$ GeV.  In order to check our background 
estimates, we used continuum 
data collected just
below the $\Upsilon$(1S) resonance, center-of-mass energy 
$9.431$ GeV $< E_{\mathrm CM} < 9.434$ GeV, 
below the $\Upsilon$(2S) resonance, 
center-of-mass energy $9.996$ GeV $< E_{\mathrm CM} < 10.004$ GeV, 
below the 
$\Upsilon$(3S) resonance, center-of-mass energy $10.329$ GeV $< E_{\mathrm CM} < 10.331$ GeV and below the $\Upsilon$(4S) resonance, center-of-mass energy 
$10.41$ GeV $< E_{\mathrm CM} < 10.57$.

To obtain a clean sample of hadronic events, we selected those events
that had a minimum of four high-quality 
charged tracks (to suppress contamination
from QED events), a total visible energy greater than 15\% of the total
center-of-mass energy (to reduce contamination from two-photon events
and beam-gas interactions), and an event vertex position 
consistent with the nominal $e^+e^-$ collision point to within
$\pm 5$ cm along the $e^+e^-$ axis (${\hat z}$) and $\pm$2 cm in the transverse
($r-\phi$) plane.
We additionally veto events with a well-defined electron or muon, or 
consistent with a $\tau\tau$ ``1-prong vs. 3-prong''
charged-track topology.
Our full data sample is summarized
in Table I. 

\begin{table}[htpb]
\label{tab:data}
\begin{tabular}{c|c|c|c|c|c}  \hline
DataSet & Resonance & ${\cal L}$ (${\rm pb}^{-1}$) & HadEvts & $\sigma^{had}_{obs}$ (nb) & EvtSel (raw) \\ 
\hline 

1S-A & $\Upsilon$(1S) & 6.351  & 128019 & 20.16 &  226746 \\

1S-B & $\Upsilon$(1S) & 633.399  & 12803279 & 20.21 &  15720815 \\

1S-C & $\Upsilon$(1S) & 424.668  & 8742850 & 20.59 &  10553140 \\

2S-A & $\Upsilon$(2S) & 450.907  & 4165745 & 9.24 & 6561803 \\

2S-B & $\Upsilon$(2S) & 6.133  & 55834 & 9.10 &  76208\\

2S-C & $\Upsilon$(2S) & 199.665  &  1839390 & 9.21 &  2748240\\

2S-D & $\Upsilon$(2S) & 248.473  & 2299910 & 9.26 &  2914640\\

2S-E & $\Upsilon$(2S) & 283.890  & 2629250 & 9.26 &  3473320\\

3S-A & $\Upsilon$(3S) & 382.902  & 2482170 & 6.52 &  3887570\\

3S-B & $\Upsilon$(3S) & 607.122  & 3948690 & 6.50 &  5736980\\

3S-C & $\Upsilon$(3S) & 180.758  & 1168980 & 6.47 &  2108220\\

1S-CO-A & $<\Upsilon$(1S) & 141.808  & 485790 & 3.43 &  619060\\

1S-CO-B & $<\Upsilon$(1S) & 46.600  & 159959 & 3.43 &  260599\\

2S-CO-A & $<\Upsilon$(2S) & 153.367  & 472071 & 3.08 &  624505\\

2S-CO-B & $<\Upsilon$(2S) & 106.409  & 326371 & 3.07 &  465939\\

2S-CO-C & $<\Upsilon$(2S) & 32.153  & 99377 & 3.09 & 138898\\

2S-CO-D & $<\Upsilon$(2S) & 59.783  & 183897 & 3.08 &  256185\\

2S-CO-E & $<\Upsilon$(2S) & 44.635  & 137083 & 3.07 &  191205\\

3S-CO-A & $<\Upsilon$(3S) & 46.906  & 135069 & 2.88 &  193749\\

3S-CO-B & $<\Upsilon$(3S) & 78.947  & 226700 & 2.87 &  321169\\

3S-CO-C & $<\Upsilon$(3S) & 32.064  & 91997 & 2.87 &  130021\\

4S-CO-A & $<\Upsilon$(4S) & 215.604  & 594662 & 2.76 & 847875 \\ 

4S-CO-B & $<\Upsilon$(4S) & 558.442  & 1536020 & 2.75 &  2189720\\ 

4S-CO-C & $<\Upsilon$(4S) & 270.896  & 753418 & 2.78 & 1073410 \\ 

4S-CO-D & $<\Upsilon$(4S) & 656.261  & 1815920 & 2.77 &  2587650\\ 

4S-CO-E & $<\Upsilon$(4S) & 238.903  & 660883 & 2.77 &  941162\\ 

4S-CO-F & $<\Upsilon$(4S) & 338.620  & 938454 & 2.77 &  1337990\\ 
\hline
\end{tabular} 
\caption{Summary of data used in analysis. 
Different running periods are 
designated by capital roman letters. For
each data set, we track the number of photons per unit
luminosity, as well as the total number of observed 
hadronic events per unit luminosity; consistency of our
results across datasets is later used as part of our
systematic error assessment. EvtSel denotes events analyzed, 
HadEvts denotes the total number of events in each sample identified as hadronic by our event selection requirements, and $\sigma$ is the corresponding 
observed hadronic cross-section for each data sample.}
\end{table}

\section*{Determination of $N_{gg\gamma}$}
\begin{figure}[htpb]
\centerline{\includegraphics[width=8cm]{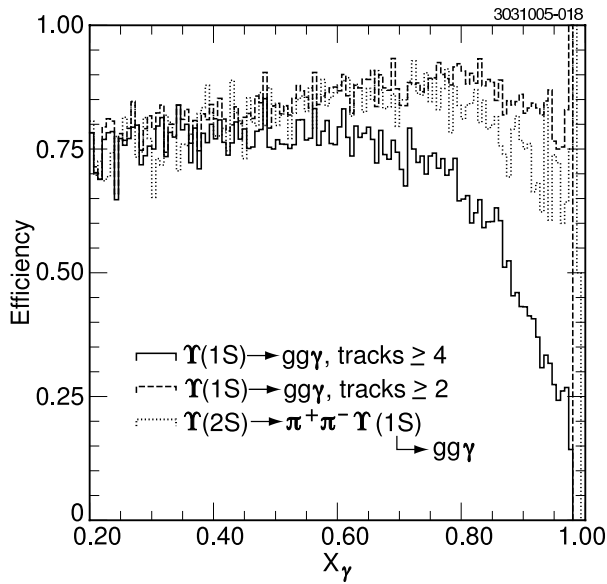}}
\caption{\label{fig:eff}Efficiency for an event containing a
fiducially contained direct photon to pass 
both event selection and shower selection
requirements for
$\Upsilon$(1S)$\to\gamma gg$ using our default photon
selection requirements and our default charged multiplicity
requirement ($\ge$4 charged tracks observed in a candidate event;
solid); $\Upsilon$(1S)$\to\gamma gg$ showing the efficiency if
the multiplicity requirement was relaxed to $\ge$2 charged tracks (dashed); 
$\Upsilon$(1S)
direct-photon daughters, for $\Upsilon$(1S) produced in $\Upsilon$(2S)
dipion decays (dotted, and
discussed later in this document). Efficiencies are derived from
full GEANT-based CLEO~III detector simulations.}
\end{figure}

To obtain $N_{gg\gamma}$, we had to determine the number of direct photon
events. Then, with the number of three-gluon events, we can extract
the ratio $R_\gamma$. For $N_{gg\gamma}$, 
only photons from
the barrel region ($|\cos\theta_z| < 0.7$) were considered.
Photon candidates were
required to be well-separated from charged tracks and other photon candidates,
with a
lateral shower shape consistent with that expected from
a true photon. 
Photons produced in the decay
of a highly energetic $\pi^0$ would sometimes produce overlapping showers in
the calorimeter, creating a so-called `merged' $\pi^0$. Two selection 
requirements
were imposed to remove this
background. First, any two photons which both have energies greater than
50 MeV and also have an opening angle 
$\theta_{\gamma_1\gamma_2}$ such that 
$\cos\theta_{\gamma_1\gamma_2}>$0.975 are removed from candidacy as direct 
photons. Second,
an effective invariant mass was determined from the energy
distribution within a single electromagnetic shower. Showers with effective
invariant masses consistent with those from merged $\pi^0$'s were also
rejected. 
After all photon and event selection requirements, the momentum-dependent
direct-photon finding efficiency is shown in Figure \ref{fig:eff},
as calculated from a large-statistics sample of photon
showers simulated with the standard, GEANT-based CLEO~III
detector simulation.
We note that, since the minimum charged multiplicity
requirement dominates the efficiency near the upper endpoint,
the $\Upsilon$(2S)$\to gg\gamma$ and
$\Upsilon$(3S)$\to gg\gamma$ direct-photon finding
efficiencies $\epsilon(x_\gamma)$ are higher than
those shown for the $\Upsilon$(1S), given their
higher initial center-of-mass energies. Our final branching
fraction calculations explicitly correct for this 
photon momentum dependence.

\begin{figure}[htpb]
\centerline{\includegraphics[width=8cm]{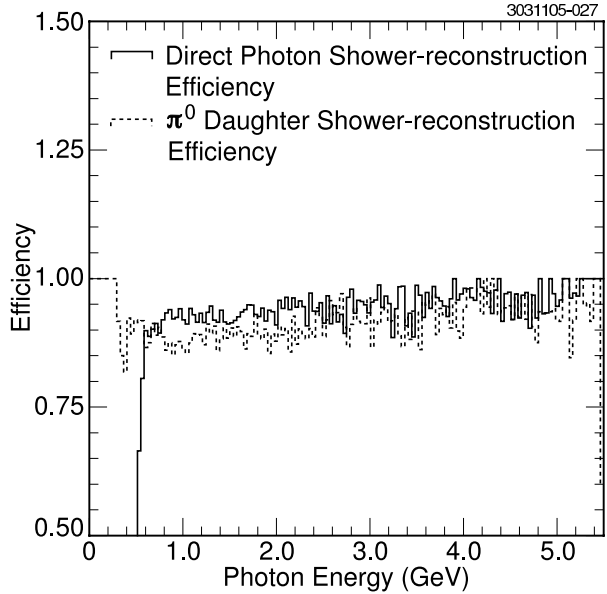}}
\caption{Comparison of shower-reconstruction
efficiency for direct photons, compared to well-separated
photons resulting from $\pi^0$ decays, based on
full GEANT-based CLEO~III simulations. 
The former efficiency is used to determine final direct
photon signal branching fractions in the data itself;
the latter is used to determine the fraction of
generated pseudo-photons which are expected to contribute to the
background showers observed in data. The difference between the two
is attributed to the typically greater isolation of direct photons.}
\label{fig:pi0gameff}
\end{figure}

Using GEANT-based CLEO~III detector simulations, 
we have compared the shower-reconstruction efficiency 
(not imposing event selection requirements) for direct
photons with the shower-reconstruction efficiency for well-separated
photons produced in the decay $\pi^0\to\gamma\gamma$; these efficiencies
are observed to agree to within 3\% over the momentum region of interest (Fig.
\ref{fig:pi0gameff}). The photon spectrum inferred from the
observed charged pion spectrum is ``multiplied'' by the dashed line in
this figure to estimate the background photons expected from
the decay of neutral pions produced in gluon and quark fragmentation.

The dominant backgrounds to the direct photon
measurement are of two types: initial
state radiation ($x_{\gamma}>0.65$) and the
overwhelming number of
background photons primarily from asymmetric
$\pi^0$ decays ($x_\gamma<$0.65), that result in two, spatially
well-separated daughter photons which elude the $\pi^0$ suppression
described above.
If our Monte Carlo generator were sufficiently accurate, of course, we
could use the 
GEANT-based 
CLEO~III Monte Carlo simulation itself to
directly generate the expected
background to the direct photon signal, including all background sources.
We have compared this GEANT-derived
photon spectrum (based on the JETSET 7.4
event generator) with data
for continuum events at $E_{\mathrm {CM}}=10.55$~GeV.
We observe fair, but not excellent agreement between the two, motivating
a data-driven estimate of the background to the direct photon signal.
We use GEANT to model the response 
of the calorimeter to photon showers (Fig.
\ref{fig:eff}), but use the data itself
as an event generator of three-gluon decays, in place of JETSET.
To model the production of $\pi^0$ daughter photons,
we took advantage of the similar kinematic distributions 
expected between charged and
neutral pions, as dictated by isospin invariance. Although isospin
conservation
will break down at low center-of-mass energies (where, e.g., the neutral
vs. charged pion mass differences and contributions from weak decays may become important), at the 
high-energy end of the spectrum (provided there is sufficient phase
space), 
we expect isospin conservation to be reliable, so 
that 
there should be half as many neutral pions as charged pions. We stress
here that this is true for three-gluon decays, $\chi_{bJ}$ decays,
${\it I}=0$ continuum $q{\overline{q}}$ events, 
${\it I}=0$ $e^+e^-\to\Upsilon\to\gamma^\star\to q{\overline q}$ 
events, etc.

There are, nevertheless, both `physics' as well as detector biases which
comprise corrections to our isospin assumption, as follows.
For continuum production
of hadrons via $\gamma^\star\to (u{\overline u}+d{\overline d})$, 
the ratio of ${\it I}=1/{\it I}=0$ production is expected to be 9:1. 
Particles with ${\it I}$=0
($\omega$, $f_0$, etc.) should decay in 
accordance with our naive assumption that $\pi^0/\pi^{\pm}$=1/2.
For sufficiently high-multiplicity decays, such that
all $\rho$ states are populated evenly, we again
expect $\pi^0/\pi^{\pm}$=1/2; very close to the threshold turn-on,
phase space effects will favor $\rho^0$ production, in which
case $\pi^0/\pi^{\pm}<$1/2. Our explicit subtraction
of the photon spectrum obtained on the continuum will
remove any such biases from 
$e^+e^-\to\gamma^\star\to q{\overline{q}}$,
leaving 
three-gluon decays as the primary background source, 
which are presumed to obey isospin conservation.

\begin{figure}[htpb]
\centerline{\includegraphics[width=8cm]{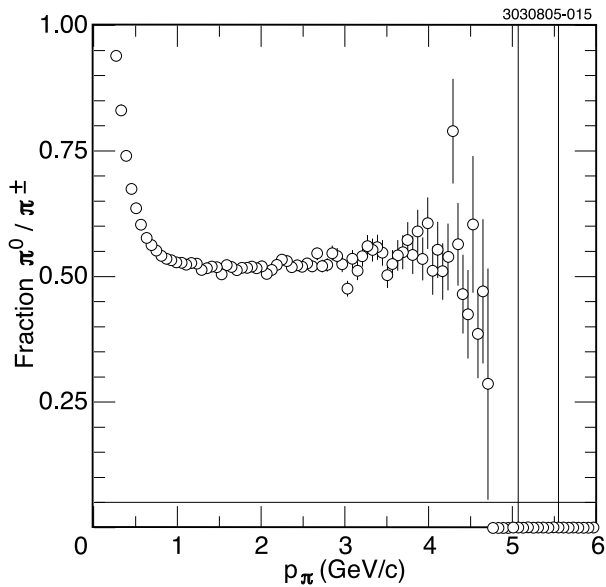}}
\caption{$\pi^0/({\pi^+}+{\pi^-})$ ratio, as a function of charged pion momentum, including tracking efficiency, particle identification efficiency and event selection requirements (from GEANT-based
CLEO~III
Monte Carlo simulations). The loss of 
efficiency at large $p_\pi$ is largely due to the 
bias introduced by the minimum 
charged-particle multiplicity 
requirement.}\label{fig:pi0picharged.eps}\end{figure}

Acceptance-related biases, which will affect both
continuum as well as resonance decays, include:
a) slight inefficiencies in our charged $\pi^\pm$
identification and tracking, b) charged kaons
and protons which fake charged pions, and c) for low multiplicity events,
an enhanced likelihood that an event with charged pions will pass our
minimum charged-multiplicity requirement compared to 
an event with neutral pions.
The $\pi^0/\pi^\pm$ ratio
therefore deviates slightly from 0.5, as a function of momentum.
Figure \ref{fig:pi0picharged.eps} 
shows the (JETSET+GEANT)-based
neutral to
charged pion production ratio for continuum 
$e^+e^-\to q{\overline q}(\gamma)$ events taking into
account such selection biases; we observe agreement
with the 0.5 expectation to within 3\%. 
For this study, we rely on JETSET 7.4 to produce the proper ratio of 
$\pi^0:\pi^\pm$ at the generator level in hadronic
$\Upsilon$ fragmentation, if not the individual spectra
themselves.
In our analysis, we use
this ratio, rather than the simple isospin
expectation, to generate pseudo-$\pi^0$'s using data charged pions as
input.
The deviation between this value and the simple isospin expectation is later
incorporated into the overall
systematic error.

\begin{figure}[htpb]
\centerline{\includegraphics[width=8cm]{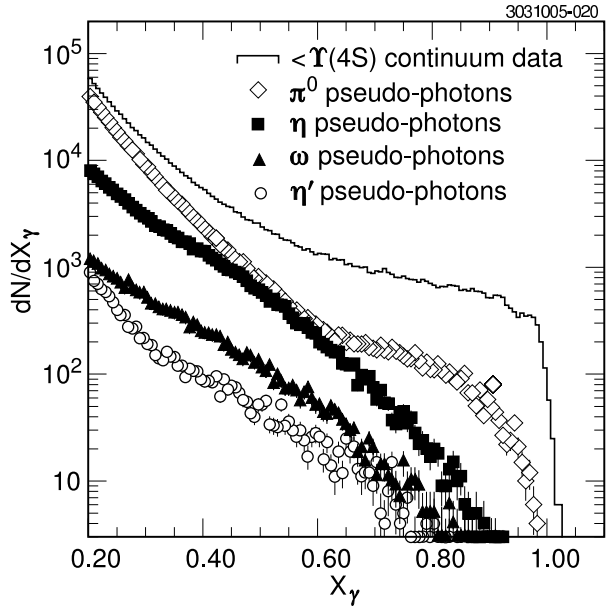}}
\caption{Estimate of the momentum-dependent contribution from various
background sources to the observed below-$\Upsilon$(4S)
inclusive photon spectrum, based on the
JETSET 7.4 event generator plus a full CLEO~III GEANT-based
detector simulation. Initial state radiation contributions are not
included.}
\label{fig:dkpiggwithcomps}
\end{figure}

These pseudo-$\pi^0$'s are 
subsequently decayed according to a 
phase space model, and the resulting simulated
photon spectrum 
is then plotted. 
It includes our GEANT-derived photon efficiency,
and the correlation between daughter photon momenta and
the photon emission direction relative to the $\pi^0$ flight direction in
the lab.
In addition to $\pi^0$'s, we also simulate
$\eta\rightarrow \gamma \gamma$, $\omega \rightarrow \pi^0 \gamma$, and
$\eta' \rightarrow \gamma (\rho, \omega, \gamma)$ contributions, using
previous measurements of these backgrounds in $\Upsilon$(1S) 
decays\cite{r:nedpaper}. An estimate of the relative contribution of
these various backgrounds to the observed 
continuum spectrum is shown in Figure
\ref{fig:dkpiggwithcomps}.

\begin{figure}[htpb]
\centerline{\includegraphics[width=8cm]{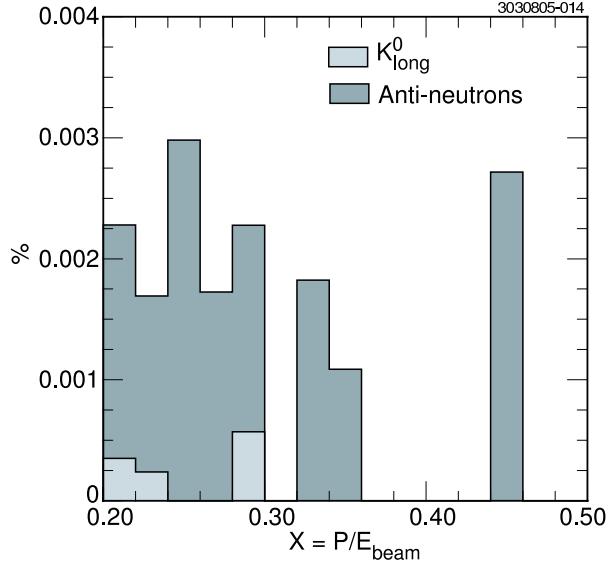}}
\caption{Estimate of the momentum dependent percentage of showers produced by 
$\bar{n}^0$'s and $K^0_L$'s that passed our shower
selection, based on a sample
of 1 million MC continuum events. There are no entries for
$x>0.5$.}
\label{fig:hadronic-contamination}
\end{figure}

We have also studied the relative contribution to the inclusive spectrum from
neutrons, anti-neutrons, and $K^0_L$'s. 
According to Monte Carlo simulations, the expected
numbers of 
such particles per hadronic event with $|\cos(\theta_z)|<0.7$, and scaled
momentum $x>$ 0.25 ({\it i.e.}, 
particles which could populate our signal region) 
are quite small.
Figure \ref{fig:hadronic-contamination} gives the yield per event, 
as a function of momentum, for $K^0_L$ and anti-neutrons
to contaminate our signal region.
Such
contributions are therefore neglected in
the remainder of the analysis.

\begin{figure}[htpb]
\centerline{\includegraphics[width=8cm]{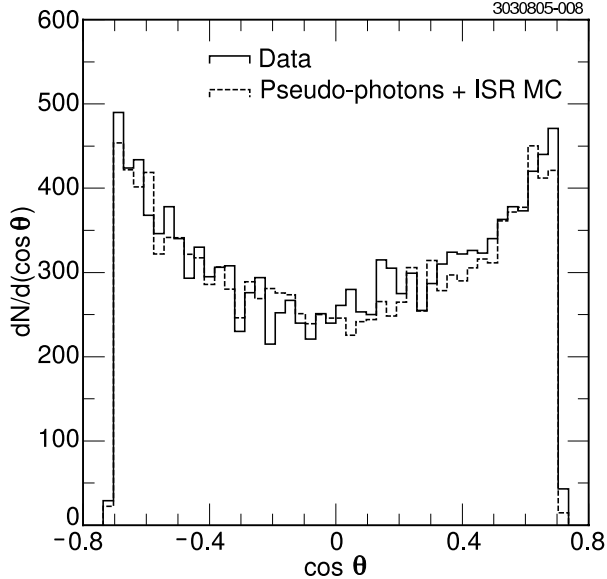}}
\caption{Angular distribution of the inclusive photons from continuum data compared with our pseudo-photon estimate, based on isospin invariance,
for showers with $x_\gamma>0.45$, and including a Monte Carlo-based estimate of the ISR background. The
normalization is absolute.}\label{fig:dkpvcon-ang}\end{figure}

\begin{figure}[htpb]
\centerline{\includegraphics[width=8cm]{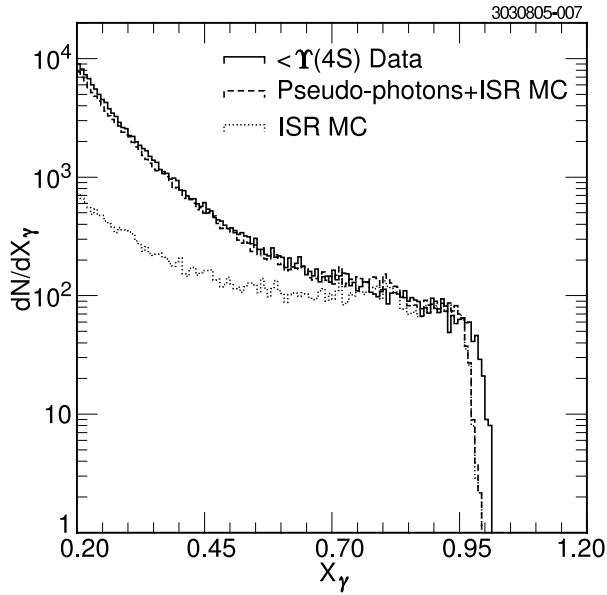}}
\caption{Comparison of $x_\gamma$ spectra,
obtained using below-$\Upsilon$(4S) data, with the sum of ISR Monte
Carlo simulations plus a pseudo-photon spectrum obtained using
identified data charged pions as input.}
\label{fig:4SCO-dkpigg-check.ps}
\end{figure}

\begin{figure}[htpb]
\centerline{\includegraphics[width=8cm]{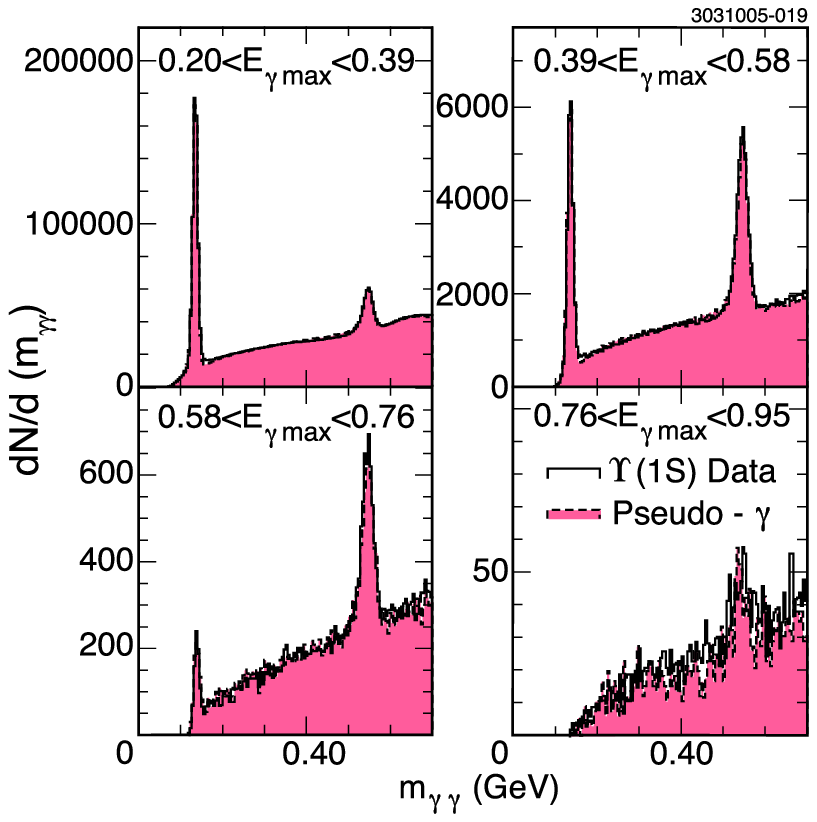}}
\caption{The $\pi^0$ and $\eta$ yields for data (dashed) and 
simulated photons (solid). The yields agree at the 2\%-3\% level.}
\label{fig:pi0noangcut.eps}
\end{figure}

The performance of our photon-background estimator can be calibrated from
data itself. Three cross-checks are presented below: a) comparison
between the absolutely normalized angular distribution of our simulated
pseudo-photons (``PP'') using
continuum charged tracks as input to our
pseudo-photon generator\footnote{Note 
that there are two simulations referred to
in this document -- ``simulated'' PP photons refer to the pseudo-photons
generated using identified charged pion tracks as inputs; ``Monte Carlo''
refers to the full JETSET+GEANT CLEO~III event+detector simulation.} 
versus the photon spectrum measured on 
the continuum (including a Monte Carlo-estimated
initial state radiation (``ISR'') contribution (Figure
\ref{fig:dkpvcon-ang})),
b) comparison of the absolute magnitude of the
pseudo-photon momentum spectrum with continuum data 
(Figure \ref{fig:4SCO-dkpigg-check.ps}),\footnote{Note that 
the Monte Carlo ISR (MC ISR) 
spectrum shows an enhancement 
in the interval $0.7<x_\gamma<0.8$, compared to the lack of events in
the region 
$0.8<x_\gamma$. This is attributable to
a) the $e^+e^-\to c{\overline c}\gamma$
threshold being crossed for $x_\gamma>0.8$, 
b) since the $e^+e^-\to q{\overline{q}}$ 
cross-section $\sigma_{q{\overline q}}\sim 1/s \sim 1/x^2_\gamma$, there
is an enhancement in hadronic final-state production as the 
energy of the radiated ISR
photon approaches the beam energy.} 
 and c) comparison of the
reconstructed $\pi^0$ and $\eta$ mass peaks (Figure \ref{fig:pi0noangcut.eps})
between our simulated photons and real data photons.
All these checks show acceptable agreement between simulation
and data. The numerical accuracy of our background estimate can
be assessed by comparing, for the second of these checks, the
fractional excess remaining after the estimated pseudo-photon
background (+ISR) is subtracted from the
raw continuum data spectrum, in the momentum interval of interest:
$(N_{\mathrm{data}}-N_{\mathrm{pseudo-photons}}-N_{\rm ISR})/N_{\mathrm{data}}$. 
Integrated from
$x_\gamma$=0.4 to $x_\gamma$=0.95, we find the fractional excesses to be
-1.86\%, -0.68\%, 2.55\% and 1.76\%, using data below the 
$\Upsilon$(1S), $\Upsilon$(2S), 
$\Upsilon$(3S) and $\Upsilon$(4S) resonances, respectively; we
consider these excesses to be acceptably consistent with zero.

Other systematic checks of our data (photon yield per data set, 
comparison between the below-$\Upsilon$(1S),
below-$\Upsilon$(2S), below-$\Upsilon$(3S) and below-$\Upsilon$(4S)
continuum photon momentum spectra) indicate good internal consistency
of all data sets considered.

\subsection*{Signal Extraction}
Two different methods were used to subtract 
background photons
and obtain 
the $\Upsilon$(1S)$\to gg\gamma$, 
$\Upsilon$(2S)$\to gg\gamma$ and 
$\Upsilon$(3S)$\to gg\gamma$ spectra.
In the first, 
after explicitly subtracting the continuum
photon spectrum from data taken on-resonance, 
we use the pseudo-photon spectrum to model the
background due to $\pi^0$, 
$\eta$, $\eta'$ and $\omega$ decay which must be separated from 
direct photons from $\Upsilon$ decay. 
In the second method, 
we used an exponential parametrization of the background
to estimate the non-direct photon contribution.

\begin{figure}[htpb]
\centerline{\includegraphics[width=8cm]{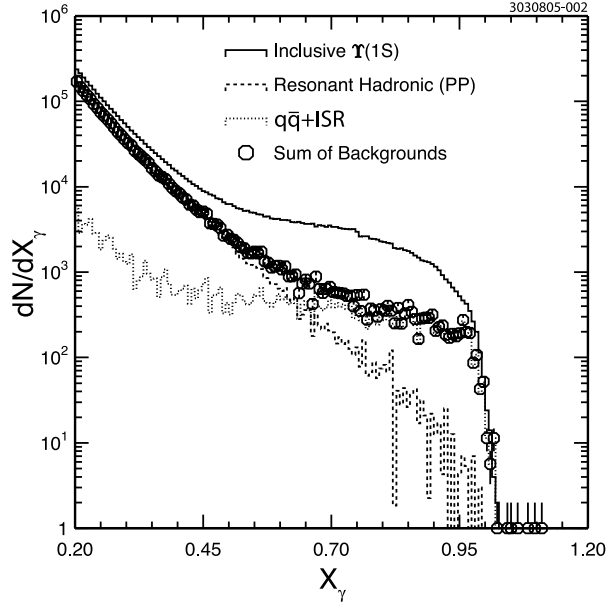}}
\caption{Photon energy spectrum (minimal cuts) for data taken at the
$\Upsilon$(1S) resonance energy, with continuum 
contribution (using CLEO data taken off the resonance, including ISR) 
and $\Upsilon$ non-direct simulated
pseudo-photons (``pp'') resulting from decays of neutral hadrons overlaid.}
\label{fig:1sinc+bkgd}
\end{figure}

\begin{figure}[htpb]
\centerline{\includegraphics[width=8cm]{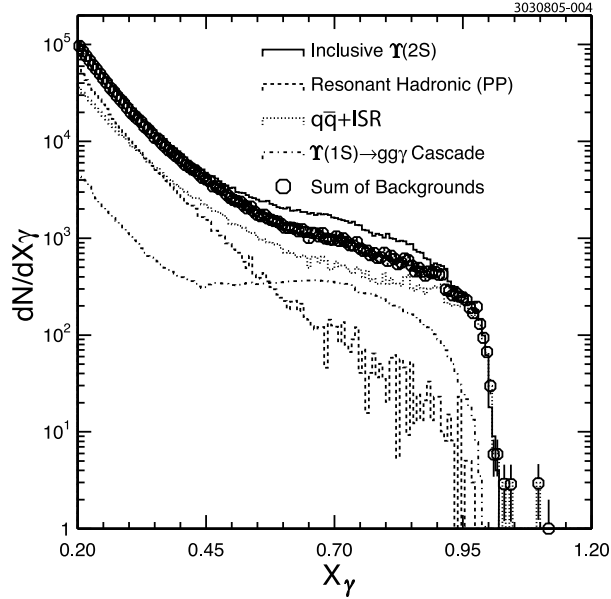}}
\caption{$\Upsilon$(2S) photon energy spectrum (minimal cuts), with 
non-direct pseudo-photons, continuum background photons, and
the cascade contribution from $\Upsilon$(1S) decays overlaid.}
\label{fig:2sinc+bkgd}
\end{figure}

\begin{figure}[htpb]
\centerline{\includegraphics[width=8cm]{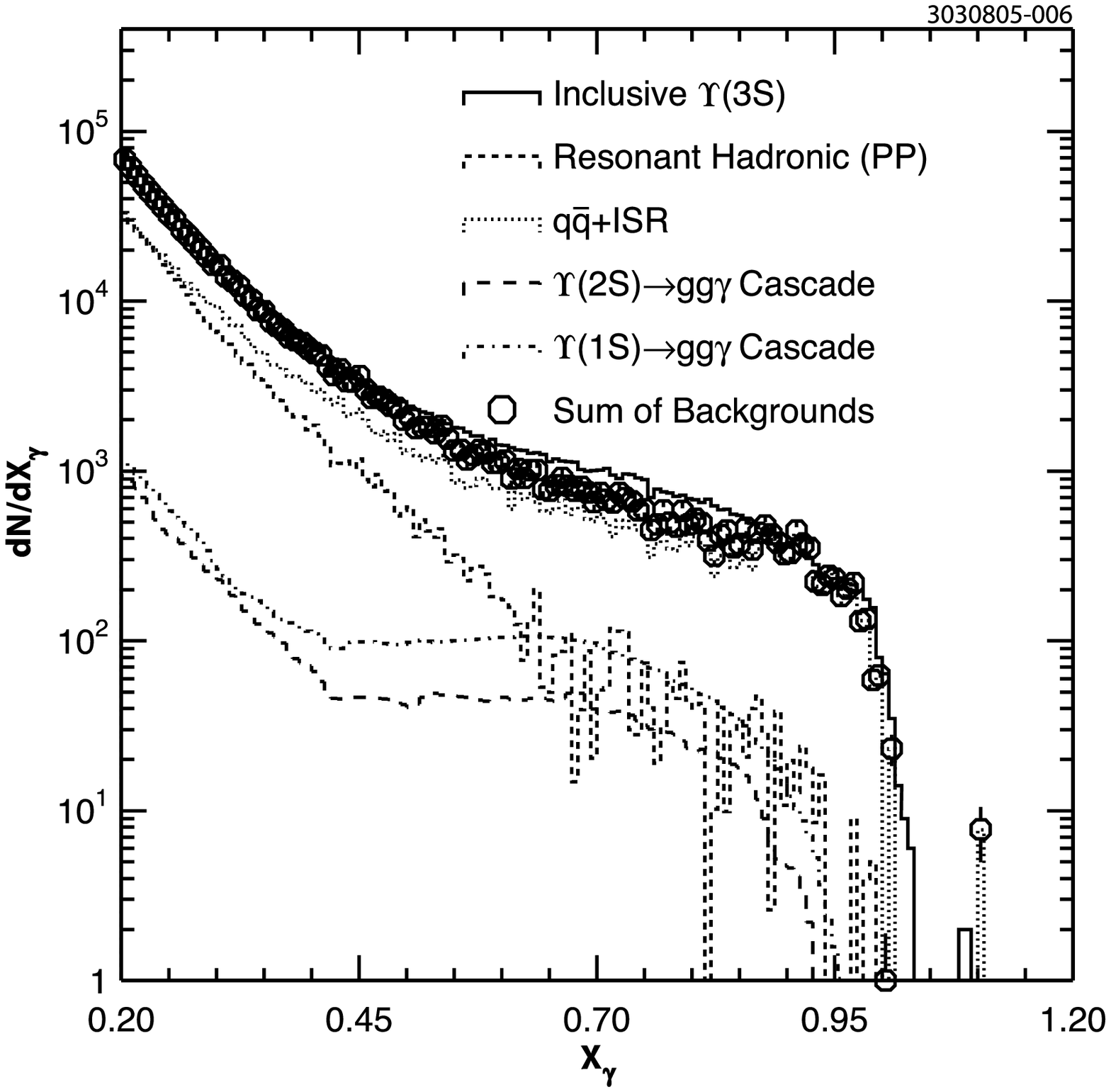}}
\caption{$\Upsilon$(3S) photon energy spectrum (minimal cuts), with
non-direct pseudo-photons, continuum background photons, 
and the cascade contributions from $\Upsilon$(2S) and $\Upsilon$(1S) decays
overlaid.}
\label{fig:3sinc+bkgd}
\end{figure}

Figure~\ref{fig:1sinc+bkgd} shows the inclusive $\Upsilon$(1S) photon distribution with
the different
estimated background contributions (continuum
photons from all sources and $\Upsilon$
decays of neutral hadrons into photons) 
overlaid.  After subtracting these sources, what remained
of the inclusive $\Upsilon$(1S) spectrum was identified as 
the direct photon spectrum,
$\Upsilon$(1S)$\rightarrow gg\gamma$.
Figures \ref{fig:2sinc+bkgd} and \ref{fig:3sinc+bkgd} show the 
corresponding plots for the $\Upsilon$(2S) and 
$\Upsilon$(3S) data, and also indicate the magnitude of the cascade
subtraction due to transitions of the type $\Upsilon$(2S)$\to\Upsilon$(1S)+X,
$\Upsilon$(1S)$\to\gamma gg$. We use currently tabulated values for
$\Upsilon$(2S)$\to\Upsilon$(1S)+X to determine the magnitude of this
correction.  Monte Carlo simulations of the primary cascade processes,
including $\Upsilon$(2S)$\to\pi\pi\Upsilon$(1S) (using a Yan\cite{r:Yan}
distribution for the dipion mass distribution) and
$\Upsilon$(2S)$\to\chi_b\gamma$, $\chi_b\to\gamma\Upsilon$(1S) are
used to adjust the shape of our measured $\Upsilon$(1S)$\to\gamma gg$
direct photon spectrum to
that expected for the cascade subtraction in order to account
for the shifted kinematic endpoint and Doppler smearing of the
daughter $\Upsilon$(1S) direct photon spectrum.
We assume that the daughter $\Upsilon$(1S) retains the 
polarization of the parent $\Upsilon$(2S); the direct
photon angular distribution is then the same as for 
direct production and
decay of the $\Upsilon$(1S) resonance.

\subsection*{Parametric estimate of background}
\begin{figure}[htpb]
\centerline{\includegraphics[width=8cm]{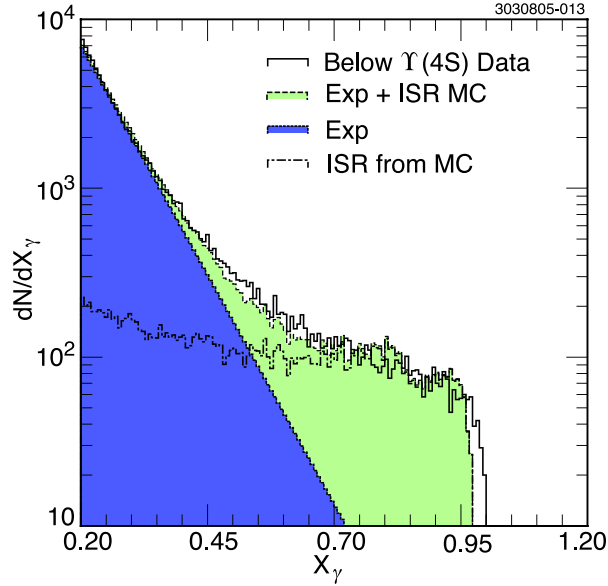}}
\caption{Subtraction of backgrounds using an exponential
(below-4S continuum data), with floating
normalization to estimate the non-direct photon spectrum.
The exponential, plus initial
state radiation, gives a fair match to the observed spectrum, although
the background is clearly underestimated in the intermediate region of the
momentum spectrum.}
\label{fig:exp4sCO.ps}
\end{figure}

Observing that the photon spectrum seems to describe
an exponential outside the signal region, 
we attempted to check our pseudo-photon and continuum-subtracted yields
against the signal photon yields
obtained when we simply fit the background to an exponential in the 
momentum region below the signal region (comparing the results obtained from
fitting $0.2<x_\gamma<0.3$ to those obtained using
$0.3<x_\gamma<0.4$)
and then extrapolated to the region $0.4<x_\gamma$. 
Figure \ref{fig:exp4sCO.ps} shows that this procedure satisfactorily
reproduces continuum data below the $\Upsilon$(4S) resonance, verifying
that it may be used to generate a rough estimate of the backgrounds.

\subsection*{Model Fits}
\begin{figure}[htpb]
\centerline{\includegraphics[width=8cm]{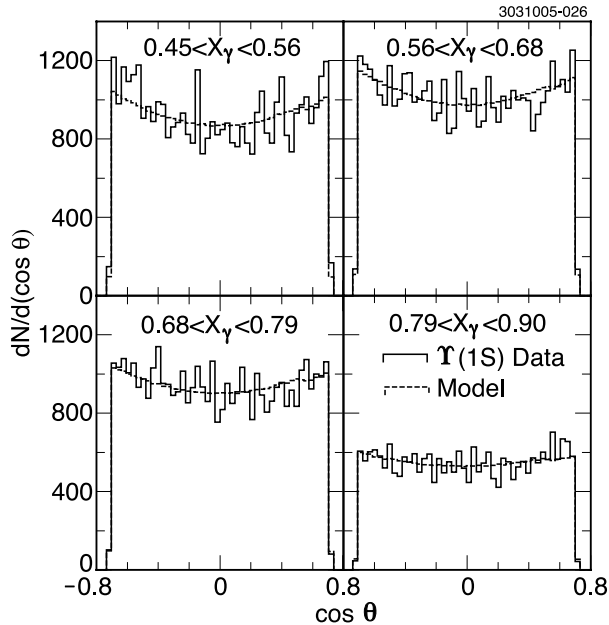}}
\caption{Photon angular distribution for background-subtracted direct
photon data (histogram) vs. Koller-Walsh prediction, modified for
the experimental efficiency as a function of $x_\gamma$ and
$\cos\theta_z$.} 
\label{fig:angular_model_check.eps}
\end{figure}

We estimate $R_\gamma$ by extrapolating the 
background-subtracted photon spectrum down
to $x_\gamma$=0, using a model
to prescribe the spectral shape at low photon
momentum. Since the CLEO calorimeter
has finite resolution, and since the
photon-finding efficiency is momentum-dependent, two procedures may be
used to compare with models. Either a migration-matrix can be determined
from Monte Carlo simulations to estimate the bin-to-bin smearing, with
a matrix-unfolding technique used to compare with prediction, or
the model can first be efficiency-attenuated (as a function of
momentum) and then smeared by the
experimental resolution to compare with data. We have followed
the latter procedure, floating only the normalization of the
efficiency-attenuated, resolution-smeared model, in this analysis. 
To determine the percentage of direct photons within our fiducial acceptance,
we used the QCD predictions of Koller and Walsh 
for the direct-photon energy and
angular distributions\cite{r:kol-walsh}.
Our large statistics sample allows (for the first time) a
check of the Koller-Walsh prediction. Figure \ref{fig:angular_model_check.eps}
shows that the angular distribution of our data, after 
taking into account acceptance effects, agrees adequately with the
Koller-Walsh prediction.

\begin{figure}[htpb]
\centerline{\includegraphics[width=8cm]{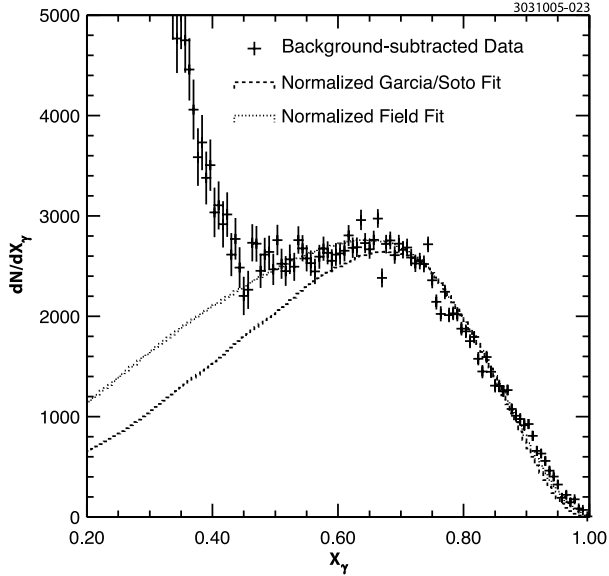}}
\caption{Fit to background-subtracted 
$\Upsilon$(1S) data, using explicit continuum data subtraction.
Garcia-Soto model is used for spectral shape
(modified for efficiency and experimental resolution), either
using a $\chi^2$ fit in the region where the direct photon contribution
dominates, or normalizing the model to the experimental data in the
same interval, as shown. The two fits
very nearly overlay with each other.}
\label{fig:1sco-gsfit}
\end{figure}

\begin{figure}[htpb]
\centerline{\includegraphics[width=8cm]{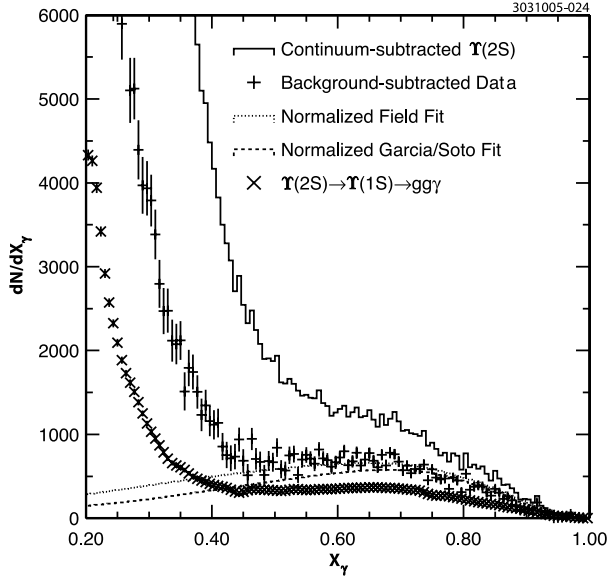}}
\caption{Fit to background-subtracted 
$\Upsilon$(2S) data, using explicit continuum data subtraction and
explicit subtraction of $\Upsilon$(1S) cascade contributions.
Direct spectrum fit using Garcia-Soto model.}
\label{fig:2sco-gsfit}
\end{figure}

\begin{figure}[htpb]
\centerline{\includegraphics[width=8cm]{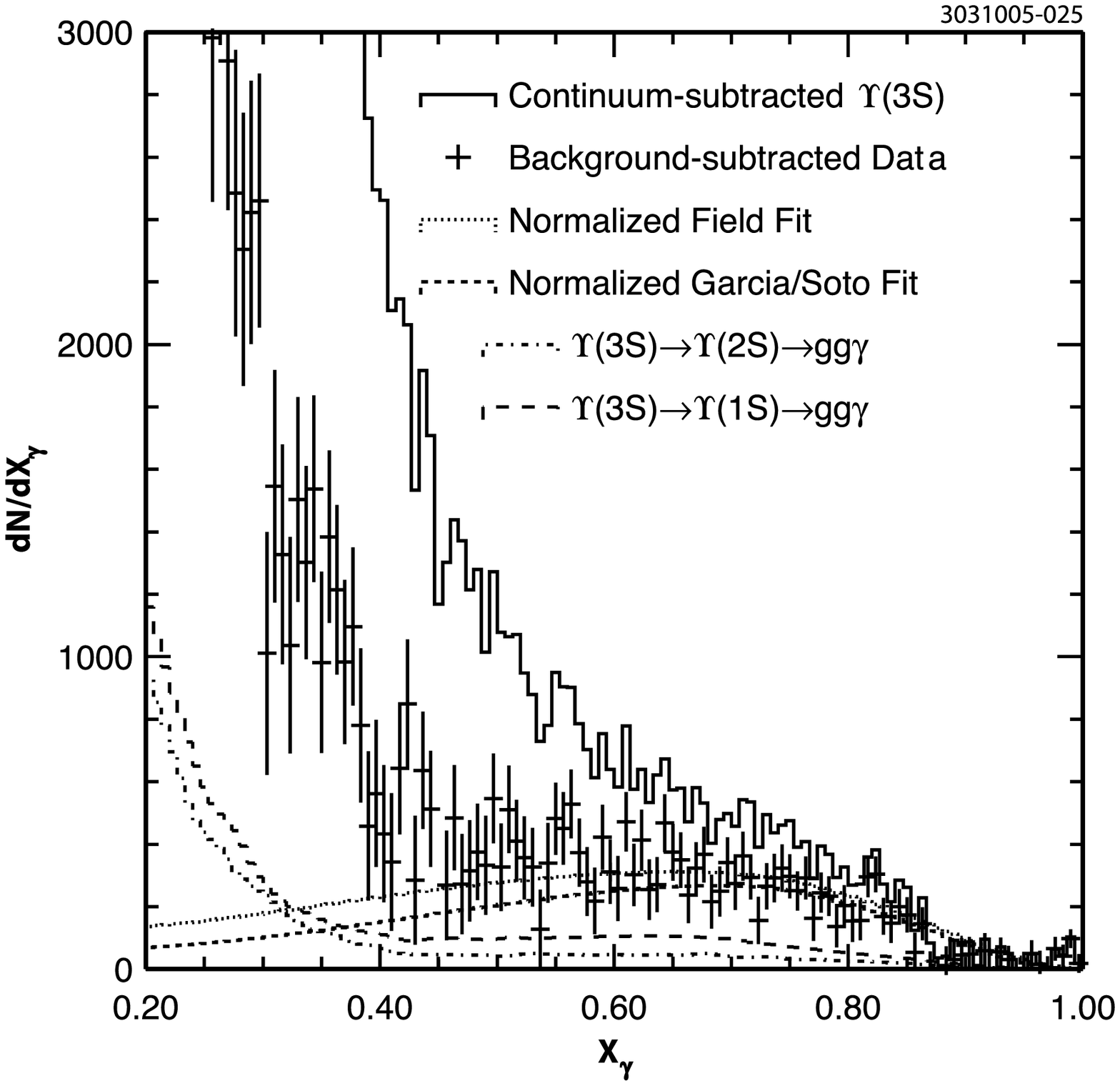}}
\caption{Fit to background-subtracted 
$\Upsilon$(3S) data, using explicit continuum data subtraction and
explicit subtraction of $\Upsilon$(1S) and $\Upsilon$(2S) cascade 
contributions.} 
\label{fig:3sco-gsfit}
\end{figure}

Figures \ref{fig:1sco-gsfit},
\ref{fig:2sco-gsfit}, and \ref{fig:3sco-gsfit} show the fits
of the direct photon energy spectrum to the
Garcia-Soto direct photon
model. The fits are performed over the interval
claimed to be relatively free of either endpoint effects or
fragmentation backgrounds (0.65$\le x_\gamma\le$0.92), then extrapolated
under these backgrounds into the unfit
region using only the direct photon component of their spectral model.
Field prescribes no
such cut-offs, so we have fit that model over the larger
kinematic range 0.4$<x_\gamma<$0.95. To probe fitting
systematics, we have performed two fits. In the first, we perform a
simple $\chi^2$ minimization of the background-subtracted data to
the Garcia-Soto spectrum. In the second, we have normalized the
area of the theoretical spectrum to the area of the background-subtracted
data in the interval of interest. The two methods yield nearly identical
results.

We note, in some cases, an excess of photons in
data as $x_\gamma\to1$. Further examination of these events indicate that
they are dominated by
$e^+e^-\to\gamma\pi^+\pi^-\pi^+\pi^-$. 

\begin{figure}[htpb]
\centerline{\includegraphics[width=8cm]{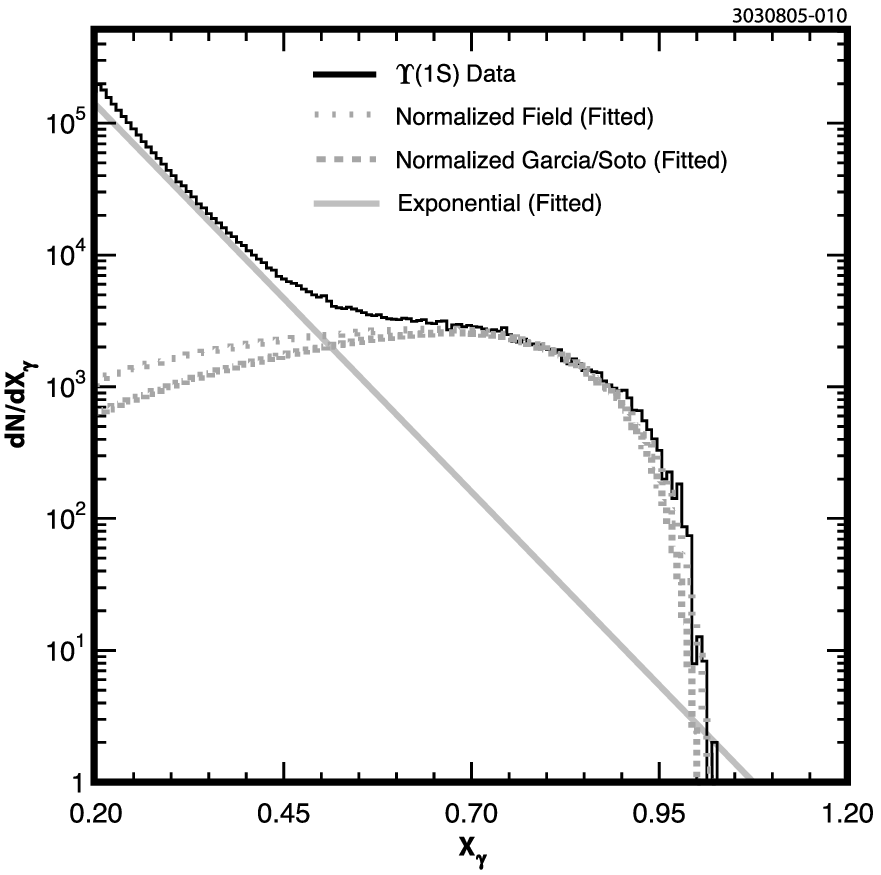}}
\caption{Subtraction of backgrounds using an exponential
($\Upsilon$(1S) data), with floating
normalization to estimate the non-direct photon spectrum.  Direct spectrum fit using Field model.}
\label{fig:exp1s-fieldfit}
\end{figure}

\begin{figure}[htpb]
\centerline{\includegraphics[width=8cm]{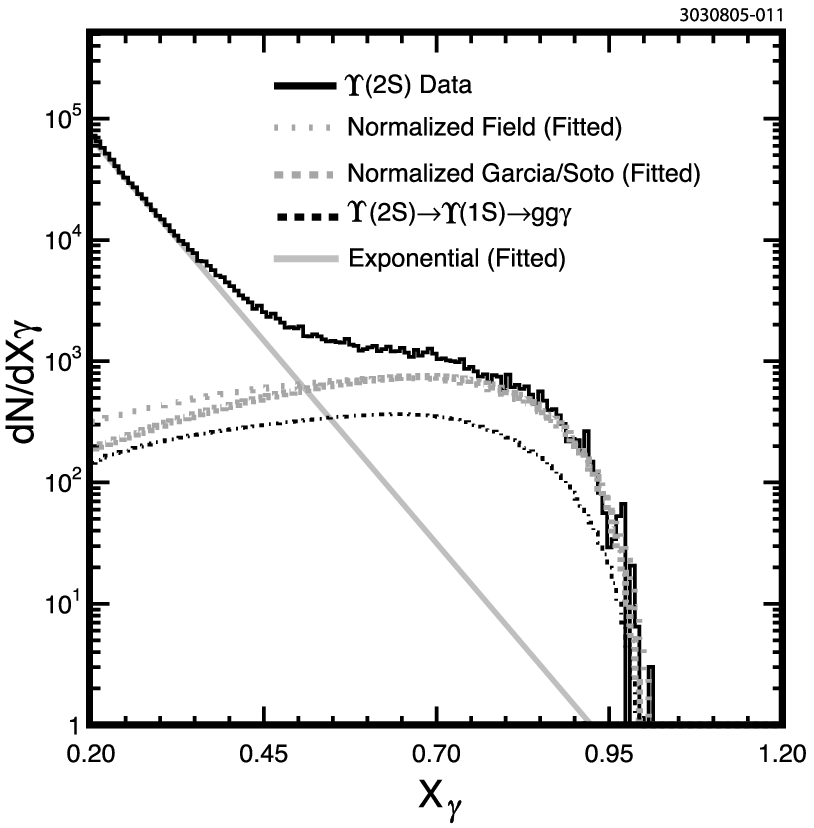}}
\caption{Subtraction of backgrounds using an exponential
($\Upsilon$(2S) data), with floating
normalization to estimate the non-direct photon spectrum.  Direct spectrum fit using Field.}
\label{fig:exp2s-fieldfit}
\end{figure}

\begin{figure}[htpb]
\centerline{\includegraphics[width=8cm]{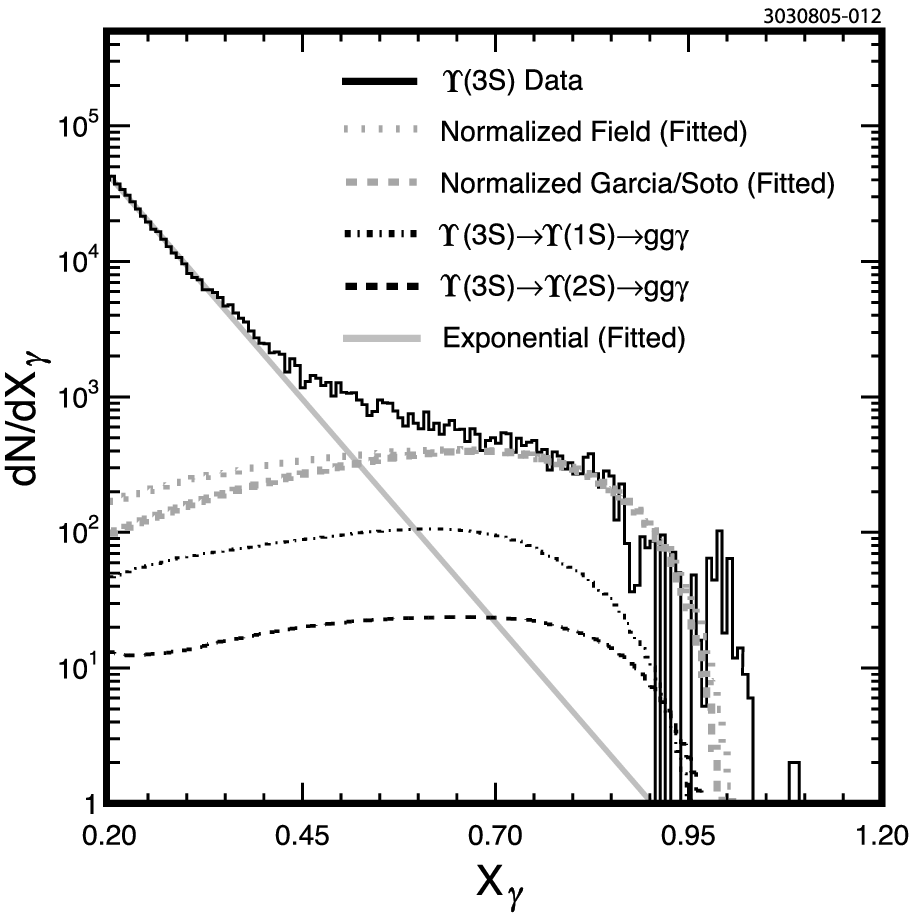}}
\caption{Subtraction of backgrounds using an exponential
($\Upsilon$(3S) data), with floating
normalization to estimate the non-direct photon spectrum.  Direct spectrum fit using Field.}
\label{fig:exp3s-fieldfit}
\end{figure}

Figures \ref{fig:exp1s-fieldfit}, \ref{fig:exp2s-fieldfit}, and
\ref{fig:exp3s-fieldfit} show fits obtained using a simple
exponential parametrization of the background, with no pseudo-photon
generation.

\subsection*{$\Upsilon$(2S)$\to\pi^+\pi^-\Upsilon$(1S); 
$\Upsilon$(1S)$\to\gamma gg$}
\begin{figure}[htpb]
\centerline{\includegraphics[width=8cm]{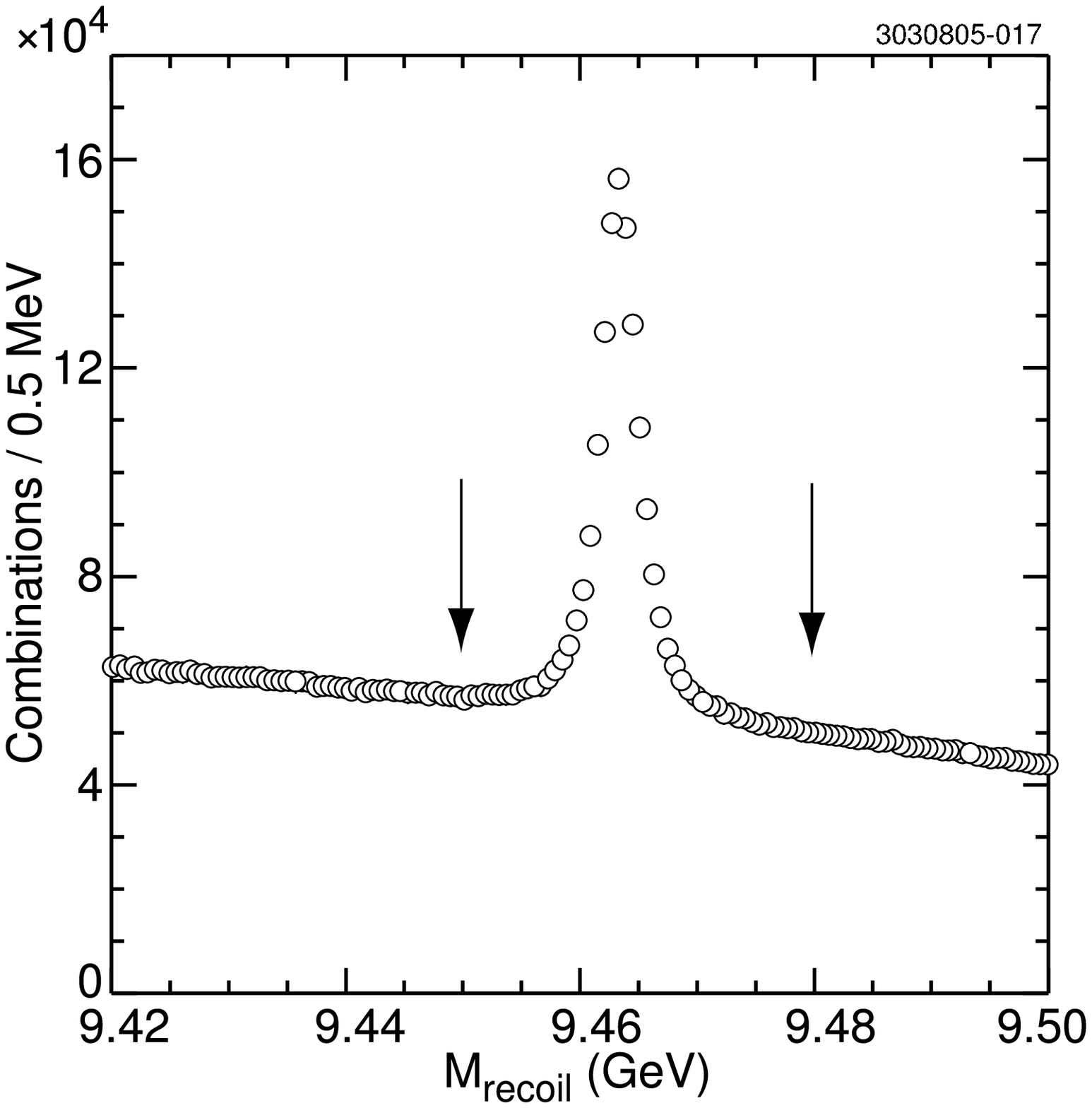}}
\caption{Mass recoiling against oppositely signed charged 
pion pairs, $\Upsilon$(2S) data.}
\label{fig:pipirecoil}
\end{figure}

Our large sample of $\Upsilon$(2S) decays
and the substantial
$\Upsilon$(2S)$\to\pi^+\pi^-\Upsilon$(1S) branching fraction ($\sim$0.19) 
afford an opportunity
to measure a `tagged' $\Upsilon$(1S) direct photon spectrum which
circumvents all continuum backgrounds. In a given event taken
at the $\Upsilon$(2S) center-of-mass energy, we calculate
the mass recoiling against all oppositely-signed charged 
pion pairs (Figure \ref{fig:pipirecoil}).
In each bin of recoil mass, we plot the spectrum of all high-energy
photons in that event. A sideband subtraction around the 
$\Upsilon$(2S)$\to\pi^+\pi^-\Upsilon$(1S) recoil mass signal, at
$m_{\rm recoil}\sim\Upsilon$(1S) results in a tagged $\Upsilon$(1S) direct
photon spectrum. Spectral shape and $R_\gamma$ values obtained this way
are consistent with our other estimates.

\section*{Determination of $N_{ggg}$}
To determine the number of three-gluon events $N_{ggg}$ from the number of
observed $\Upsilon$(1S) hadronic events $N^{\Upsilon({\rm 1S})}_{had}$, 
we first subtracted the number
of continuum events 
at the $\Upsilon$(1S) energy
$N^{\Upsilon({\rm 1S})}_{cont}$ based on the observed number of
below-$\Upsilon$(1S) continuum events $N^{E_{\mathrm CM}=9.43~GeV}_{cont}$:
\begin{equation}
N^{\Upsilon({\rm 1S})}_{cont} = N^{E_{\mathrm CM}=9.43~GeV}_{cont} \cdot
                 \frac{{\cal L}_{\Upsilon({\rm 1S})}s_{\rm cont}}{{\cal L}_{\rm cont}s_{\rm 1S}} \cdot
\end{equation}
\noindent
where the factor $\frac{ {\cal L}_{\Upsilon({\rm 1S})} } { {\cal L}_{\rm cont}}$ arises from the 
number of events $N = {\cal L} \cdot \sigma$, and $s\equiv E_{\mathrm CM}^2$.
From the observed number of hadronic events collected at the
resonance
$N^{\Upsilon({\rm 1S})}_{had}$, and knowing the 
branching fractions and efficiencies for
$\Upsilon({\rm 1S})\to q{\overline{q}}$,
$\Upsilon({\rm 1S})\to gg\gamma$, and
$\Upsilon({\rm 1S})\to ggg$, the 
number of $\Upsilon$(1S)$\to ggg$ events can be inferred.
For $\Upsilon({\rm 1S})\to q{\overline{q}}$, e.g.,
we use 
the averaged
$\Upsilon$(1S)$\rightarrow \mu^+ \mu^-$ branching fraction $B_{\mu \mu} =
0.0248$~\cite{r:pdg,r:istvan}, and $R_{\Upsilon({\rm 1S})} = \sigma(e^+e^- \rightarrow \Upsilon$(1S)$\rightarrow
q{\bar q}) / \sigma(e^+e^- \rightarrow \Upsilon$(1S)$\rightarrow \mu^+ \mu^-) =
3.51$~\cite{r:Albrecht-92,r:CLEO-R}:
\begin{equation}
N_{q{\overline q}}=
R_{\Upsilon({\rm 1S})}\cdot 
B_{\mu\mu}\epsilon_{q{\overline q}} \cdot         
\frac{N^{\Upsilon({\rm 1S})}_{had}}{(1 - 3 B_{\mu \mu}){\overline\epsilon}_{had}}
\end{equation}
\noindent
with ${\overline\epsilon}_{had}$ the hadronic event reconstruction
efficiency, averaged over all hadronic modes, and 
$\epsilon_{q{\overline q}}$ specifically the efficiency for
reconstructing an $\Upsilon{\rm (1S)}\to(q{\overline q})$ event.

Using ${\cal B}(\Upsilon$(2S)$\to\Upsilon{\rm (1S)}+X$)=(32$\pm$1)\%,
${\cal B}(\Upsilon$(3S)$\to\Upsilon{\rm (2S)}+X$)=(10.6$\pm$0.8)\%, and
${\cal B}(\Upsilon$(3S)$\to\Upsilon{\rm (3S)}+X$)=(12.1$\pm$0.5)\%, 
three-gluon
decay fractions 
for the three resonances $f_{ggg}$=$81.3\pm0.5$\%, $39\pm1$\%, and
$38\pm1$\%, respectively, 
we obtained a value for $N_{ggg}$ for each of the
resonances,
based on our measured values for $N_{gg\gamma}$.
More details on this subtraction are presented in 
Table II.

\begin{table}[htpb]
\label{tab:eventeff}
\begin{center}
\begin{tabular}{|c|c|}\hline
Event Type & Efficiency ($\epsilon$)  \\
\hline
$\Upsilon$(1S)$\rightarrow{ggg}$ & 0.953 $\pm$ 0.003 \\
$\Upsilon$(1S)$\rightarrow{gg\gamma}$ & 0.751 $\pm$ 0.007\\
$\Upsilon$(1S)$\rightarrow{q\bar{q}}$ & 0.871 $\pm$ 0.005\\
$\Upsilon$(2S)$\rightarrow{ggg}$ & 0.956 $\pm$ 0.003\\
$\Upsilon$(2S)$\rightarrow{gg\gamma}$ & 0.776 $\pm$ 0.007\\
$\Upsilon$(2S)$\rightarrow{q\bar{q}}$ & 0.882 $\pm$ 0.005 \\
$\Upsilon$(2S)$\rightarrow\Upsilon$(1S)$+X\rightarrow{ggg}$ & 0.956 $\pm$ 0.003\\
$\Upsilon$(2S)$\rightarrow\Upsilon$(1S)$+X\rightarrow{gg\gamma}$ & 0.778 $\pm$ 0.007\\
$\Upsilon$(2S)$\rightarrow\Upsilon$(1S)$+X\rightarrow{q\bar{q}}$ & 0.891 $\pm$ 0.005\\
$\Upsilon$(2S)$\rightarrow\chi_{bJ}$(1P)$\rightarrow{gg}$ $(J=0,1,2)$ & 0.933 $\pm$ 0.004\\
$\Upsilon$(3S)$\rightarrow{ggg}$ & 0.955 $\pm$ 0.003\\
$\Upsilon$(3S)$\rightarrow{gg\gamma}$ & 0.765 $\pm$ 0.007\\
$\Upsilon$(3S)$\rightarrow{q\bar{q}}$ & 0.881 $\pm$ 0.005\\
$\Upsilon$(3S)$\rightarrow\Upsilon$(2S)$+X\rightarrow{ggg}$ & 0.958 $\pm$ 0.003\\
$\Upsilon$(3S)$\rightarrow\Upsilon$(2S)$+X\rightarrow{gg\gamma}$ & 0.765 $\pm$ 0.007\\
$\Upsilon$(3S)$\rightarrow\Upsilon$(2S)$+X\rightarrow{q\bar{q}}$ & 0.877 $\pm$ 0.005\\
$\Upsilon$(3S)$\rightarrow\Upsilon$(1S)$+X\rightarrow{ggg}$ & 0.961 $\pm$ 0.003\\
$\Upsilon$(3S)$\rightarrow\Upsilon$(1S)$+X\rightarrow{gg\gamma}$ & 0.789 $\pm$ 0.006\\
$\Upsilon$(3S)$\rightarrow\Upsilon$(1S)$+X\rightarrow{q\bar{q}}$ & 0.90 $\pm$ 0.07\\
$\Upsilon$(3S)$\rightarrow\chi_{bJ}$(1P)$\rightarrow{gg}$ $(J=0,1,2)$ & 0.819 $\pm$ 0.006\\
$\Upsilon$(3S)$\rightarrow\chi_{bJ}$(2P)$\rightarrow{gg}$ $(J=0,1,2)$ & 0.929 $\pm$ 0.004\\
\hline
$\Upsilon$ Resonance & $N_{\rm total}(\Upsilon$(nS)) $(10^6)$ \\
\hline
$\Upsilon$(1S) & 21.0 $\pm$ 0.06\\
$\Upsilon$(2S) & 8.4 $\pm$ 0.04\\
$\Upsilon$(3S) & 5.2 $\pm$ 0.06\\
\hline
Fraction & $f$ \\
\hline
$f(\Upsilon$(1S)$\rightarrow{ggg})$ & 0.813 $\pm$ 0.005 \\
$f(\Upsilon$(2S)$\rightarrow{ggg})$ & 0.39 $\pm$ 0.01 \\
$f(\Upsilon$(3S)$\rightarrow{ggg})$ & 0.38 $\pm$ 0.01 \\
$f(\Upsilon$(2S)$\rightarrow\Upsilon$(1S)$+X)$ & 0.32 $\pm$ 0.01 \\
$f(\Upsilon$(3S)$\rightarrow\Upsilon$(2S)$+X)$ & 0.106 $\pm$ 0.008 \\
$f(\Upsilon$(3S)$\rightarrow\Upsilon$(1S)$+X)$ & 0.121 $\pm$ 0.005 \\
\hline
\end{tabular}
\end{center}
\caption{Efficiencies for the reconstruction of the 
various types of events considered in this analysis, the total number of 
calculated $\Upsilon$(1S), 
$\Upsilon$(2S) and $\Upsilon$(3S) events,
and the fractions of these totals which were used to obtain 
the $\Upsilon\rightarrow{ggg}$ denominator in our measurements 
of $R_\gamma$ (and to scale the direct photon cascade spectra in our 
$\Upsilon$(2S) and 
$\Upsilon$(3S) subtractions).  
These fractions were obtained from the Particle Data Group\cite{r:pdg} and 
include recent CLEO $\chi_{bJ}$ measurements\cite{r:chib}.  
The presented errors on the efficiencies are statistical only. Note that,
although there are more possible decay paths from the $\Upsilon$(3S) than
the $\Upsilon$(2S), the $ggg$ fractions are comparable owing to the 
significantly larger $\Upsilon$(2S)$\to\Upsilon$(1S)$\pi\pi$ branching
fraction.}
\end{table}

\section*{Results}
With values of $N_{gg\gamma}$ and $N_{ggg}$, the
ratio $R_\gamma$ can be determined.
Table III 
presents our numerical results for the
extracted branching fractions.
\begin{table}[htpb]
\label{tab:Results-sum}
\begin{tabular}{c|c|c|c} \hline
$X\to gg\gamma$; $X$= & Background & Field $R_\gamma$ / $\chi^2$/d.o.f. 
& GS $R_\gamma$ / $\chi^2$/d.o.f. \\ \hline
$\Upsilon$(1S) & Exponential  & $(2.94\pm0.02)$\% / 115.1/67-1 & $(2.39\pm0.03)$\% / 132.4/67-1 \\
$\Upsilon$(1S) & PP (MC ISR)  & $(2.81\pm0.01)$\% / 293/74-1 & $(2.48\pm0.01)$\% / 694/74-1 \\
$\Upsilon$(1S) & PP (CO ISR)  & $(2.93\pm0.01)$\% / 125/67-1  & $(2.45\pm0.01)$\% / 116/37-1\\
$\Upsilon({\rm 2S})\to\pi\pi\Upsilon({\rm 1S})$-tagged & PP (no ISR)  & $(2.9\pm0.3)$\% / 118/58-1 & $(2.5\pm0.3)$\% / 132/58-1 \\
$\Upsilon$(2S) & Exponential  & $(3.7\pm0.7)$\% / 542/105-1 & $(3.4\pm0.4)$\% / 773/105-1\\
$\Upsilon$(2S) & PP (MC ISR)  & $(3.42\pm0.05)$\% / 316/67-1 & $(3.01\pm0.04)$\% 426/67-1 \\
$\Upsilon$(2S) & PP (CO ISR)  & $(3.58\pm0.05)$\% 145/67-1 & $(2.77\pm0.05)$\% 87/37-1 \\
$\Upsilon$(3S) & Exponential  & $(3.4\pm0.4)$\% 210/105-1 & $(3.1\pm0.1)$\% 251/105-1\\
$\Upsilon$(3S) & PP (MC ISR)  & $(2.91\pm0.07)$\% / 263/67-1 & $(2.55\pm0.06)$\% 331/67-1 \\
$\Upsilon$(3S) & PP (CO ISR)  & $(2.8\pm0.1)$\% / 72/67-1 & $(2.1\pm0.1)$\% / 36/37-1 \\
\end{tabular} 
\caption{Summary of Measurements. ``PP'' denotes Pseudo-Photon background,
``MC ISR'' implies that Monte Carlo simulations of initial state radiation 
were used to subtract the ISR background. These numbers are provided
for comparison only and are not used in final averaging, etc.
``CO ISR'' implies that data ISR
was subtracted directly using below-resonance data. 
Values obtained using an exponential parametrization
have had systematic errors (reflecting sensitivity to region chosen for scale
normalization of the exponential outside the peak region; for this estimate, 
branching fractions were compared using the regions $0.2<x_\gamma<0.3$ or
$0.3<x_\gamma<0.4$ to set the scale of the exponential and extrapolate under
the signal in the higher-$x_\gamma$ region) added in quadrature with the statistical
error. All other errors are statistical only.
Note that $\Upsilon$(2S) $R_\gamma$ values have been corrected for $\Upsilon$(1S)$\to gg\gamma$ contamination; $\Upsilon$(3S) $R_\gamma$ values have been corrected for both $\Upsilon$(2S)$\to gg\gamma$, and also $\Upsilon$(1S)$\to gg\gamma$ contamination.}\end{table}
We note that, in general, the reduced $\chi^2$ values for the fits tend to be
rather high.
Structure in the spectrum due to, e.g., two-body radiative decays, 
may result in such a poor fit and is 
currently being investigated. 
The normalization-by-area fits probe the
extent to which the model fits
may be disproportionately weighted by a small number of points.

\section*{Systematic errors}
We identify and estimate systematic errors as follows:
\begin{enumerate}
\item For the $\Upsilon$(1S), the uncertainty in $N_{ggg}$ is based on
the CLEO estimated three-gluon event-finding efficiency
uncertainty. For the
$\Upsilon$(2S) and $\Upsilon$(3S) decays, the uncertainty in $N_{ggg}$
also
folds in uncertainties in the tabulated radiative and hadronic
transition decay rates
from the parent $\Upsilon$'s, which
are necessary for determining $N_{ggg}$ as well as the magnitude of the
cascade subtractions.
The cascade subtraction errors include 
statistical ($1\sigma$) 
uncertainties in the various
decay modes of the $\Upsilon$
resonances.
\item Background normalization and background 
shape uncertainty are evaluated redundantly as follows:
\begin{enumerate}
\item We determine the branching fractions with and without
an explicit $\pi^0$ veto on the background.
\item We measure the internal consistency of our
results using different sub-samples of our $\Upsilon$(1S), $\Upsilon$(2S) and 
$\Upsilon$(3S)
samples.
\item Bias in background subtraction 
can also be estimated using Monte Carlo simulations.
We treat the simulation as we do data, and generate pseudo-photons based
on the Monte Carlo identified charged pion tracks. After subtracting the
pseudo-photon spectrum from the full Monte Carlo photon spectrum, we can
compare our pseudo-photon and ISR-subtracted 
spectrum with the known spectrum that
was generated as input to the Monte Carlo detector simulation. 
For the $\Upsilon$(1S), $\Upsilon$(2S), and $\Upsilon$(3S), we observe
fractional deviations of +5.7\%, --3.4\%, and --2.1\% between the input
spectrum and the pseudo-photon 
background-subtracted spectrum. To the extent that the
initial state radiation estimate and the photon-finding efficiency are
obtained from the same Monte Carlo simulations, this procedure is largely
a check of our generation of the pseudo-photon background and the correlation
of $\pi^0$ decay angle with efficiency.
\item We extract the direct photon branching fractions using 
a flat $\pi^0:\pi^\pm$ isospin ratio of 0.5, compared to 
the $\pi^0:\pi^\pm$ ratio based on Monte Carlo simulations, 
including all our event
selection and charged tracking and charged particle identification
systematics ($\sim$0.53, Figure \ref{fig:pi0picharged.eps}).
\item Our uncertainty in the on-resonance vs. off-resonance
luminosity scaling, which determines the magnitude of the
continuum subtraction is assessed as $<$1\%, absolute.
\end{enumerate}
\item Model dependence of the extracted total decay rate is estimated
by: i) determining the variation between fits (for a given model)
performed using a $\chi^2$ minimization prescription, or a simple
normalization of area of theoretical spectrum to data, and also by
ii) comparing the results obtained from fits to the Field model with
results obtained from the fits to the Garcia-Soto model.
(We currently assume the Koller-Walsh prescription for the
angular distribution is correct, and assign no systematic error for
a possible corresponding uncertainty.) The
irreducible model-dependence error (the difference 
between branching fractions obtained with
the Field model vs. Garcia-Soto)
is presented as the last error in our
quoted branching fraction.

\end{enumerate}

There is currently no theoretical consensus on either the
shape or magnitude of the fragmentation photon background to
the direct photon spectrum.
We assign no explicit systematic error to the 
uncertainty in this component
and presume this to be already probed by the
variation observed between models, the consistency we observe
with results obtained from an exponential fit to the background, and
the consistency observed in fitting over different intervals of the
background-subtracted spectrum.
Given the currently tabulated upper limit on 
$\Upsilon$(1S)$\to\gamma$+pseudoscalar, 
pseudoscalar$\to h^+h^-$ (${\cal B}<3\times 10^{-5}$) and the small
branching fractions measured for other two-body exclusive radiative decays
(like the $\eta(1440)$, whose dominant modes do not have 
two charged tracks in the final state), 
we neglect distortions to the direct photon yield from 
exclusive two-body decays $\Upsilon\to\gamma+{\cal X}$. 

Table IV 
summarizes the systematic errors studied in this analysis
and their estimated effect on $R_{\gamma}$.  

\begin{table}[htpb]
\label{tab:sys}
\begin{center}
\begin{tabular}{|c|c|}\hline
Source                           &$\delta R_{\gamma} (1S/2S/3S)$  \\
\hline
Difference (MC Glevel, MC analyzed)                  & 0.08/0.05/0.03 \\
\hline Background Shape/Norm, including: &  \\
With/Without a $\pi^0$ veto           &0.07/0.07/0.07                    \\

 Isospin Assumption & 0.01/0.01/0.01 \\
 $\epsilon_{ggg}$ & 0.08/0.19/0.28 \\
Cascade subtraction & 0/0.07/0.14\\                    
\hline
 Luminosity and $\sqrt{s}$ 
scaling         & 0.01/0.01/0.01                    \\

\hline Fit systematics (norm vs. $\chi^2$ fit)	& 0.01/0.05/0.01\\ 
\hline\hline
{\bf Total Systematic Error} & 0.13/0.22/0.32 \\ 
\hline {\bf Model Dependence (GS vs. Field)} & 0.24/0.41/0.37 \\

\hline
\end{tabular}
\end{center}
\caption{Systematic Errors.} 

\end{table}

\subsection*{Comparison with previous analyses}
Table V 
compares the results of this analysis with those obtained by previous
experiments, 
in which the number of $\Upsilon$(1S)$\rightarrow gg \gamma$ events
were determined using Field's theoretical model only.

\begin{table}[htpb]
\begin{center}
\label{tab:comparison}
\begin{tabular}{|c|c|} \hline
 Experiment                 & $R_{\gamma}     (\%)$ \\
\hline
 CLEO 1.5 ($\Upsilon$(1S))\cite{r:Csorna}   & $2.54 \pm 0.18 \pm0.14$ \\
 ARGUS ($\Upsilon$(1S))\cite{r:Albrecht-87} & $3.00 \pm 0.13 \pm0.18$ \\
 Crystal Ball 
($\Upsilon$(1S))\cite{r:Bizeti}       & $2.7  \pm 0.2  \pm 0.4$ \\
 CLEO II ($\Upsilon$(1S))\cite{r:nedpaper}    
& $2.77 \pm 0.04 \pm0.15$ \\ \hline
 {\bf CLEO III ($\Upsilon$(1S))}   & {\bf $2.70\pm0.01\pm0.13\pm0.24$} \\
 {\bf CLEO III ($\Upsilon$(2S))} & {\bf $3.18\pm0.04\pm0.22\pm0.41$} \\
 {\bf CLEO III ($\Upsilon$(3S))} & {\bf $2.72\pm0.06\pm0.32\pm0.37$} \\
\hline
\end{tabular}
\end{center}
\caption{Comparison with other experiments. Errors are statistical, 
systematic, and model-dependent (Field vs. Garcia/Soto),
respectively. Central values are obtained by a direct
weighted average
(taking into account both statistical and systematic errors,
and assuming errors to be uncorrelated, therefore yielding
the most conservative estimate of the experimental precision) of the
measurements presented in Table III.}
\end{table}

\section*{Summary}
We have re-measured the 
$\Upsilon$(1S)$\to\gamma gg/\Upsilon$(1S)$\to ggg$ branching fraction 
ratio ($R_\gamma$), obtaining agreement with previous results. We also
have made first measurements of $R_\gamma$(2S) and $R_\gamma$(3S). Our
results are, within errors, consistent with the naive expectation that
$R_\gamma$(1S)$\sim R_\gamma$(2S)$\sim R_\gamma$(3S), although this 
equality does not hold for the 
recent CLEO measurements of $B_{\mu\mu}$ for the 
three $\Upsilon$ resonances\cite{r:danko}. 
Assuming an energy scale equal to the
parent $\Upsilon$ mass, our values of $R_\gamma$ for 
$\Upsilon$(1S)$\to gg\gamma$
(2.70$\pm$0.01$\pm$0.13$\pm$0.24)\%,  
$\Upsilon$(2S)$\to gg\gamma$ 
(3.18$\pm$0.04$\pm$0.22$\pm$0.41)\%,
and 
$\Upsilon$(3S)$\to gg\gamma$ (2.72$\pm$0.06$\pm$0.32$\pm$0.37)\%
imply values of the strong coupling
constant $\alpha_s(M_Z)$ = ($0.1114 \pm 0.0002 \pm 0.0029 \pm 0.0053$),
($0.1026 \pm 0.0007 \pm 0.0041 \pm 0.0077$) and
($0.113 \pm 0.001 \pm 0.007 \pm 0.008$), respectively,
which are within errors, albeit
consistently lower, compared to the current world average (Appendix I).

% CURRENT acknowledgements go here...
% download from the CLEO website 
% http://www.lns.cornell.edu/restricted/CLEO/analysis/ac_help/ack.html
% This is the current version:

\section*{Acknowledgments}
We thank Sean Fleming,
Xavier Garcia, Adam Leibovich, and Joan Soto for particularly
enlightening discussions.
We gratefully acknowledge the effort of the CESR staff
in providing us with excellent luminosity and running conditions.
D.~Cronin-Hennessy and A.~Ryd thank the A.P.~Sloan Foundation.
This work was supported by the National Science Foundation,
the U.S. Department of Energy, and
the Natural Sciences and Engineering Research Council of Canada.

\vspace{1cm}
\appendix*{\large Appendix~I) Calculation of strong coupling constant}

The decay width $\Upsilon \rightarrow gg \gamma$ has been calculated by Lepage
and Mackenzie~\cite{r:Mack-Lep} in terms of the energy involved in the decay
process ({\it i.e.}, $\alpha_s(E_{\mathrm CM})$, or $\alpha_s(M_{\Upsilon}$)):
\begin{equation}
\frac{\Gamma(\Upsilon \rightarrow gg \gamma)}
     {\Gamma(\Upsilon \rightarrow \mu^+\mu^-)}
     = \frac{8(\pi^2 - 9)}{9 \pi \alpha_{QED}} \alpha_s^2(M_{\Upsilon})
       \Biggl[1 + (3.7\pm0.4)\frac{\alpha_s(M_{\Upsilon})}{\pi} \Biggr].
       \label{eq:qcd0}
\end{equation}
\noindent
Sanghera~\cite{r:Sanghera} rewrites this 
expression in terms of an arbitrary energy (renormalization) scale
$\mu$:
\begin{equation}
 \frac{\Gamma(\Upsilon \rightarrow gg \gamma)}
      {\Gamma(\Upsilon \rightarrow \mu^+\mu^-)}
      = A_{\gamma}\biggl({\alpha_s(\mu)\over \pi}\biggr)^2 +
        A_{\gamma}\biggl({\alpha_s(\mu)\over \pi}\biggr)^3
                  \biggl[2 \pi b_0
                         \ln\biggl({\mu^2 \over M_{\Upsilon}^2}\biggr) +
                         (3.7 \pm 0.4)
                   \biggr], \label{eq:qcd1}
\end{equation}
\noindent
where $ A_{\gamma} = {8\pi (\pi^2-9) \over 9\alpha_{QED}}$,
$b_0=(33-2n_f)/12\pi$, and $n_f$ is the number of light quark flavors
which participate in the process ($n_f=4$ for $\Upsilon$(1S) decays).

Similarly, the decay width $\Upsilon \rightarrow ggg$ has been calculated
by Bardeen {\it et al.}~\cite{r:Bardeen} and expressed by Lepage
{\it et al.}~\cite{r:Brod-Lep-Mack,r:Mack-Lep-PRL} as:
\begin{equation}
 \frac{\Gamma(\Upsilon \rightarrow ggg)}
      {\Gamma(\Upsilon \rightarrow \mu^+\mu^-)}
      = \frac{10(\pi^2 - 9)}{81 \pi e_b^2}
        \frac{\alpha_s^3(M_{\Upsilon})}{\alpha^2_{QED}}
        \Biggl[1 + \frac{\alpha_s(M_{\Upsilon})}{\pi}
              [(2.770\pm0.007)\beta_0 - (14.0\pm0.5)] + \cdot \cdot \cdot \Biggr]
        \label{eq:qcd101}
\end{equation}
with $\beta_0 = 11 - ({2 \over 3})n_f$, and $e_b = -{1 \over 3}$, the charge
of the $b$ quark.  
Here again Sanghera~\cite{r:Sanghera} uses the same algebraic
technique to rewrite this in terms of the renormalization scale:
\begin{equation}
 \frac{\Gamma(\Upsilon \rightarrow ggg)}
      {\Gamma(\Upsilon \rightarrow \mu^+\mu^-)}
      = A_g\biggl({\alpha_s(\mu)\over \pi}\biggr)^3 +
        A_g\biggl({\alpha_s(\mu)\over \pi}\biggr)^4
           \biggl[3 \pi b_0 \ln\biggl({\mu^2 \over M_{\Upsilon}^2}\biggr) -
           \biggl({2 \over 3}\biggr)B_f n_f + B_i \biggr] \label{eq:qcd2}
\end{equation}
\noindent
with $ A_g = {10\pi^2(\pi^2-9) \over 81 e_b^2} {1 \over \alpha_{QED}^2}$,
$ B_f = 2.770 \pm 0.007$, and $B_i = 16.47 \pm 0.58$.

Note that the scale dependent QCD equations~(\ref{eq:qcd1})
and~(\ref{eq:qcd2}) are finite order in $\alpha_s$.  
If these equations were solved
to all orders, then they could in principle be used to determine $R_{\gamma}$
independent of the renormalization scale.  But since we are dealing with
calculations that are finite order, the question of an appropriate scale
value must be addressed.

The renormalization scale may be defined in terms of the center of mass
energy of the process, $\mu^2= f_\mu E_{\mathrm CM}^2$, where $f_\mu$ is some positive
fraction. Since QCD does not tell us {\it a priori} what $f_\mu$ should be,
we must define the appropriate scale.  One
possibility would be to define $\mu=E_{\mathrm CM}$; that is $f_\mu$=1. 
A number of
prescriptions~\cite{r:Brod-Lep-Mack,r:Grunberg,r:Stevenson,r:Sanghera} have
been proposed in an attempt to ``optimize'' the scale.  However, each of these
prescriptions yields scale values which, in general, vary greatly with the
experimental quantity being measured~\cite{r:Sanghera}.  
We have chosen $f_\mu=1$
to facilitate a calculation of $\alpha_s$ 
at each of the $\Upsilon$ resonance energies.

For the $\Upsilon$(1S) analysis, using $\mu = M_{\Upsilon(1S)}$ we find 
\begin{equation}
\alpha_s(M_{\Upsilon(1S)}) = 0.1735 \pm 0.0005 \pm 0.0072 \pm 0.0133;
\end{equation}
\noindent

for the $\Upsilon$(2S) analysis, using $\mu = M_{\Upsilon(2S)}$ we find 
\begin{equation}
\alpha_s(M_{\Upsilon(2S)}) = 0.151 \pm 0.002 \pm 0.009 \pm 0.017;
\end{equation}
\noindent

for the $\Upsilon$(3S) analysis, using $\mu = M_{\Upsilon(3S)}$ we find 
\begin{equation}
\alpha_s(M_{\Upsilon(3S)}) = 0.172 \pm 0.003 \pm 0.018 \pm 0.021.
\end{equation}
\noindent

The errors are statistical, systematic and model-dependent, respectively.  
These calculations 
were obtained by finding the zeroes of the ratio of 
Eqs.~\ref{eq:qcd1} and 
\ref{eq:qcd2} given 
our measurement of $R_{\gamma}$ for each $\Upsilon$
resonance.
The errors were obtained by shifting our measurement of
$R_{\gamma}$ by $\pm\sigma$, for each of
our three errors, and extracting $\alpha_s$ for each 
relevant error-shifted 
central value.

These results can then be extrapolated to $\mu = M_{Z}$ 
using equation~\ref{eq:othermu}
\cite{r:pdg} with $\mu_0 = M_{\Upsilon}$ for each resonance.  
For this calculation,
only the first three terms of
the $\beta$-function were considered\cite{r:ritt}.

\begin{equation}
\log{\frac{\mu^2}{\mu_0^2}}=\int^{\alpha_s(\mu)}_{\alpha_s(\mu_0)}{\frac{d\alpha}{\beta(\alpha)}}
\label{eq:othermu}
\end{equation}

This calculation for 
the $\Upsilon$(1S), $\Upsilon$(2S) and $\Upsilon$(3S) results in the
following measurements of $\alpha_s(M_Z,\Upsilon({\rm nS}))$:

\begin{equation}
\alpha_s^{M_Z,\Upsilon({\rm 1S})} = 0.1114 \pm 0.0002 \pm 0.0029 \pm 0.0053,
\end{equation}
\noindent

\begin{equation}
\alpha_s^{M_Z,\Upsilon({\rm 2S})} = 0.1026 \pm 0.0007 \pm 0.0041 \pm 0.0077,
\end{equation}
\noindent

\begin{equation}
\alpha_s^{M_Z,\Upsilon({\rm 3S})} = 0.113 \pm 0.001 \pm 0.007 \pm 0.008.
\end{equation}
\noindent

Our results are systematically low compared with 
the average value of $\alpha_s({M_Z}) = 0.119 \pm 0.006$
obtained from many variables studied at all the LEP
experiments~\cite{r:pdg}, but in
better agreement with $\alpha_s({M_Z}) = 0.112 \pm 0.003$ obtained
from an analysis of structure functions in deep inelastic
scattering~\cite{r:Benvenuti} and with the previous CLEO measurement of 
$\alpha_s(M_Z,\Upsilon({\rm 1S}))$ \cite{r:nedpaper}.
For the $\Upsilon$(2S) and $\Upsilon$(3S) measurements, 
we stress caution in
interpreting these results, as it is (again) unclear what 
procedure should be
used to define the renormalization scale.  

\begin{figure}[htpb]
\centerline{\includegraphics[width=9cm]{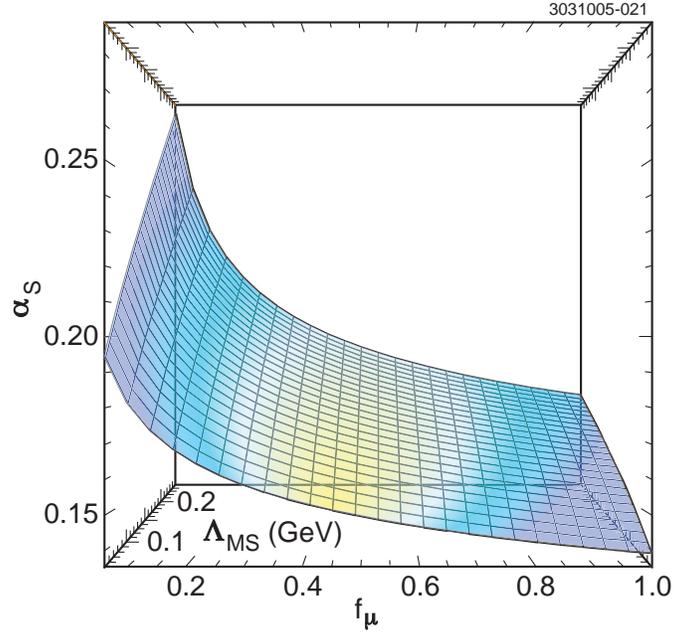}}
\caption{Contour plot illustrating dependence of $\alpha_s$ on the QCD scale
parameter $\Lambda_{\overline{MS}}$ and the momentum scale $f_\mu$.}
\label{fig:alphas-contour}
\end{figure}

As an alternative to the extraction method outlined above,
the strong coupling constant $\alpha_s$ can be written as a
function of the QCD scale parameter $\Lambda_{\overline{MS}}$, defined in
the modified minimal subtraction scheme (MMSS)\cite{r:pdg}.
Figure \ref{fig:alphas-contour} presents the contour plot of
$\alpha_s(\Lambda_{\overline{MS}},~f_\mu$).

\begin{figure}[htpb]
\centerline{\includegraphics[width=9cm]{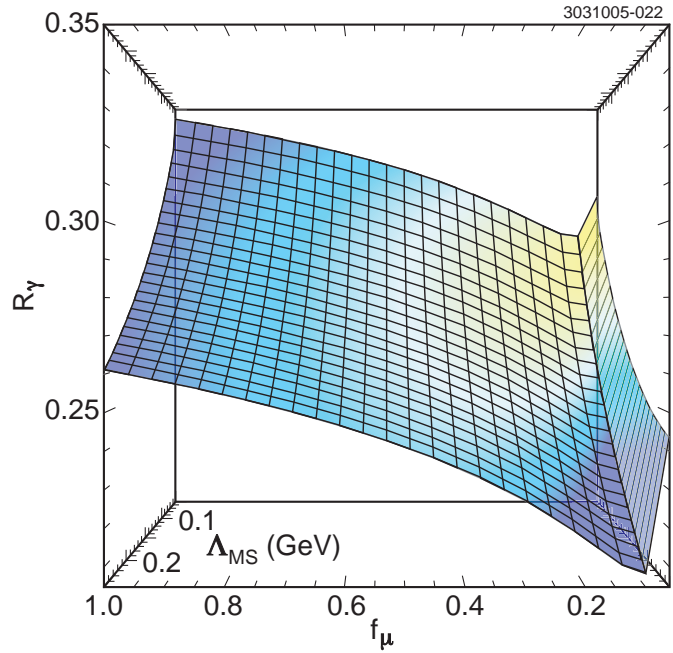}}
\caption{Contour plot illustrating relationship between $R_\gamma$, the QCD scale
parameter $\Lambda_{\overline{MS}}$ and the momentum scale $f_\mu$.}
\label{fig:rgamma-contour}
\end{figure}

Similarly, the ratio of eqns. (2) and (4) above can be used to eliminate
$\alpha_s$ and provide a relationship between
$R_\gamma$, $\Lambda_{\overline{MS}}$, and $f_\mu$ (Figure 
\ref{fig:rgamma-contour}).

\end{document}